\documentclass[12pt,preprint]{aastex}

\newcommand\Teff{$T_{\rm eff}$}
\newcommand\6{{\footnotesize VI}}
\newcommand\5{{\footnotesize V}}
\newcommand\4{{\footnotesize IV}}
\newcommand\3{{\footnotesize III}}
\newcommand\2{{\footnotesize II}}
\newcommand\1{{\footnotesize I}}
\newcommand\lam{{$\lambda$}}
\newcommand\vsini{{$v$sin$i$}}
\newcommand\kms{$\rm{km s^{-1}}$}
\newcommand\logg{log$g$}

\newcommand\pd{\phantom{$-$}}
\newcommand\p{\phantom{:}}

\newcommand\mdot{$\dot{M}$}

\newcommand{\sref}[1]{\S\ref{#1}}

\slugcomment{Accepted version}
\shorttitle{Quantitative Studies of Magellanic Cloud OB Supergiants}
\shortauthors{Evans et al.}

\begin{document}


\title{Quantitative Studies of the Far-UV, UV and Optical Spectra of 
Late O and Early B-type Supergiants in the Magellanic Clouds\altaffilmark{1}}


\author{ C. J. Evans\altaffilmark{2},
         P. A. Crowther\altaffilmark{3},
         A. W. Fullerton\altaffilmark{4,5},
         D. J. Hillier\altaffilmark{6}}

\altaffiltext{1}{Based on observations made with the
NASA-CNES-CSA Far Ultraviolet Spectroscopic Explorer. FUSE is operated
by The Johns Hopkins University under NASA contract NAS5--32985.  Also
based in part on observations collected at the European Southern
Observatory Very Large Telescope in program 67.D-0238, plus archival
data obtained with the NASA-ESA Hubble Space Telescope and the
NASA-ESA-PPARC International Ultraviolet Explorer.}

\altaffiltext{2}{Isaac Newton Group of Telescopes, 
                 Apartado de Correos 321, 
                 38700 Santa Cruz de la Palma, 
                 Canary Islands, 
                 Spain}
\altaffiltext{3}{Dept. of Physics \& Astronomy, 
                 Hounsfield Road, 
                 University of Sheffield, 
                 Sheffield S3~7RH, 
                 UK}
\altaffiltext{4}{Dept. of Physics \& Astronomy, 
                 University of Victoria,
                 P.O. Box 3055, 
                 Victoria, BC, V8W 3P6, 
                 Canada}
\altaffiltext{5}{Center for Astrophysical Sciences,
                 Dept. of Physics \& Astronomy, 
                 The Johns Hopkins University,
                 3400 N. Charles Street, 
                 Baltimore, MD 21286}
\altaffiltext{6}{Department of Physics \& Astronomy, 
                 University of Pittsburgh, 3941 O'Hara Street, PA 15260}

\begin{abstract}
We present quantitative studies of 8 late O and early B-type
supergiants in the Magellanic Clouds using far-ultraviolet $FUSE$,
ultraviolet $IUE$/$HST$ and optical VLT-UVES spectroscopy.
Temperatures, mass-loss rates and CNO abundances are obtained using
the non-LTE, spherical, line-blanketed model atmosphere code of Hillier \& Miller
(1998).  We support recent results for lower temperatures of OB-type
supergiants as a result of stellar winds and blanketing, which amounts
to $\sim$2000\,K at B0\,Ia.  In general, H$\alpha$ derived mass-loss
rates are consistent with UV and far-UV spectroscopy, although from
consideration of the S\,\4 $\lambda\lambda$1063-1073 doublet, clumped
winds are preferred over homogenous models.  AV\,235 (B0\,Iaw) is a
notable exception, which has an unusually strong H$\alpha$ profile
that is inconsistent with the other Balmer lines and UV wind
diagnostics.  We also derive CNO abundances for our sample, revealing
substantial nitrogen enrichment, with carbon and oxygen depletion. Our
results are supported by comparison with the Galactic supergiant
HD\,2905 (BC0.7\,Ia) for which near-solar CNO abundances
are obtained.  This bolsters previous suggestions that ``normal'' OB-type
supergiants exhibit atmospheric compositions indicative of partical CNO
processing.
\end{abstract}


\section{Introduction}
\label{intro}
The {\it Far~Ultraviolet~Spectroscopic~Explorer} ($FUSE$) satellite
(Moos et al. 2000) has provided an opportunity to study populations of
early-type stars in the Milky Way and Magellanic Clouds in the
900-1200\AA\ region. Spectral atlases of $FUSE$ observations of O and
early B-type stars have recently been presented by Pellerin et al.
(2002) and Walborn et al. (2002). The principal scientific motivation
for such observations is the comparison of stellar winds from
early-type stars in different metallicity ($Z$) environments. The
predicted theoretical dependence of radiatively driven winds from
massive stars is well documented, such that \mdot(Z)~$\propto Z^{0.5-0.7}$
(Kudritzki, Pauldrach \& Puls, 1987; Vink, de Koter \& Lamers, 2001);
observationally this has yet to be firmly established. The precise
sensitivity of mass-loss to metallicity for early-type stars is keenly
sought as it is a necessary ingredient for stellar evolution models
(and in turn evolutionary synthesis models).

Towards this goal, we initiated a program to obtain high-quality
optical observations with the UV-Visual Echelle Spectograph (UVES) at
the Very Large Telescope (VLT) for the Magellanic Cloud targets in our
$FUSE$ Principal Investigator programmes. We have focused particularly
on the Clouds to avoid contamination associated with the high column
densities of H$_2$ typical of sightlines to Galactic OB-type stars
(e.g., Pellerin et al. 2002) and to minimize the effect of uncertainties in
the distances to the targets on the atmospheric analysis.  At the same
time, so-called ``unified'' stellar atmosphere models are now available
which allow for the spherical extension of supergiants and treat line
blanketing by metal species in a reasonably thorough manner.

Crowther et al. (2002; hereafter Paper I) presented the first study of
the present series for four Magellanic Cloud O-type supergiants,
employing the model atmosphere code {\sc cmfgen} (Hillier \& Miller,
1998). These extreme supergiants (all were of luminosity class
Ia$^+$) were specifically chosen to test previously adopted stellar
temperatures for O-type stars, for which Fullerton et al. (2000) had
questioned the previous plane-parallel, unblanketed results
(e.g., Vacca et al. 1996). Indeed, substantially lower temperatures
were derived for these extreme supergiants.  The combination of very
strong stellar winds together with metal line blanketing led to
temperatures 15--20$\%$ lower than those from plane-parallel models
composed solely of hydrogen and helium.  Downward revisions were also
found for Galactic O-type dwarfs (Martins et al. 2002) and supergiants
in the Galactic cluster Cyg OB2 (Herrero et al. 2002).  For our second
paper we have therefore chosen to study O-type stars with luminosity
classes in the range II to Ia$^+$, to better address the question of
the stellar temperature scale.  We have also extended our sample to
include early B-type supergiants, thereby providing an overlap between
Paper I and the recent results of Trundle et al. (2004).

In Paper~I, CNO abundances of extreme O-type supergiants were found to
differ greatly from those inferred from nebular and stellar studies in
the Magellanic Clouds. Nitrogen was strongly enhanced, with carbon
(and oxygen) moderately depleted, suggestive of mixing of unprocessed
and CNO-processed material at their surfaces. These results were
supported by a companion study by Hillier et al. (2003) which used
identical techniques.  They found unprocessed CNO abundances for
AV\,69 [OC7.5\,III((f))] and partially CNO-processed material in
AV\,83 (O7\,Iaf$^+$).  Therefore, morphologically normal OB stars
appear to show evidence for moderate levels of chemical
processing. This new study allows us to examine CNO abundances in less
extreme supergiants.

The issue of clumping in early-type stars was raised in Paper~I and by
Hillier et al.  (2003).  In order to reconcile the optical H$\alpha$ and
far-UV P\,{\sc v} $\lambda\lambda$1118-28 wind diagnostics, either
clumped winds or a reduced phosphorus abundance was required.  Massa
et al. (2003) arrived at similar conclusions from a study of $FUSE$
observations of O-type stars in the LMC. The question of clumping, and
hence reduced mass-loss rates for OB-type stars, is an important one,
although the uncertain abundance of phosphorus in the interstellar medium (ISM) 
is a major limitation. In contrast, we shall suggest that S\,{\sc iv}
$\lambda\lambda$1062-73 is an analogue of P\,{\sc v}
$\lambda\lambda$1118-28 amongst late O and early B-type supergiants,
for which the ISM elemental abundance is well known (e.g., Russell \&
Dopita 1992).  This permits firmer conclusions regarding clumping in
the winds of early-type stars.

As in Paper I, the $FUSE$ data are complemented by UV spectra from the
{\it International\,Ultraviolet\,Explorer} ($IUE$) telescope and the
{\it Hubble Space Telescope} ($HST$) archives, together with optical
data from UVES.  The observational data are presented in \sref{obs},
followed by a description of our methods in \sref{methods} and results
for each of our targets stars in \sref{analysis}, except
AV~235 which is discussed separately in \sref{av235}. In
\sref{abundances} we discuss the CNO abundances derived for our
targets and also present an analysis of the Galactic BC-type
supergiant HD\,2905 to provide a test of our methods.  Finally, we
discuss the implications of our results for published temperature
calibrations and give a comparison of our observationally derived
mass-loss rates with theoretical predictions.

\section{Observations}
\label{obs}

\subsection{Optical data}
\label{obsdetails}

Basic observational parameters for the target stars are given in Table
\ref{targets1}.  High resolution optical spectra of HDE\,269050, AV\,235,
372 and 469 were obtained during 2001 September 27-29 with UVES (using
the \#2 dichroic) at the VLT.  The standard blue-arm setting
(\lam$_{\rm c}$~=~4370\,\AA) with a single 2$\times$4k EEV CCD (15$\times$15$\mu$m
pixels), gave a spectral coverage of \lam3770--4950\,\AA.  A
non-standard setting (\lam$_{\rm c}$~=~8300\,\AA) was used for the red-arm, with an
identical EEV CCD giving wavelength coverage of \lam6400--8200\,\AA.
The 2-pixel resolution in the H$\alpha$ region was 0.09\,\AA.  Further
echelle orders into the far-red (to $\sim$10000\,\AA) were also observed
simultaneously with an MIT CCD, though due to poor signal-to-noise
these data are not used here.  In addition, UVES spectra of
HDE\,269896, AV\,70, 456 and 488 were kindly provided by Dr. Lex Kaper from
observations on 2001 September 24, obtained for a study of Diffuse
Interstellar Bands (Ehrenfreund et al., 2002).  The settings for these
observations were slightly different but the overall spectral coverage
is comparable (see Table \ref{targets2}), except for AV\,456 for which
the \lam4520--4660\,\AA~region was not available.

The data were cleaned of cosmic rays, bias corrected, flat fielded and
optimally extracted in {\sc iraf}\footnote{{\sc iraf} is distributed
by the National Optical Astronomy Observatories, which are operated by
the Association of Universities for Research in Astronomy, Inc., under
cooperative agreement with the National Science Foundation.}  (v2.11).
The echelle blaze profile was removed by fitting a two-dimensional
surface to the orders and then dividing it into the data using a
routine developed by Prof. Ian Howarth (private communication) for the Starlink
program {\sc dipso}.

\begin{figure*}
\begin{center}
\includegraphics{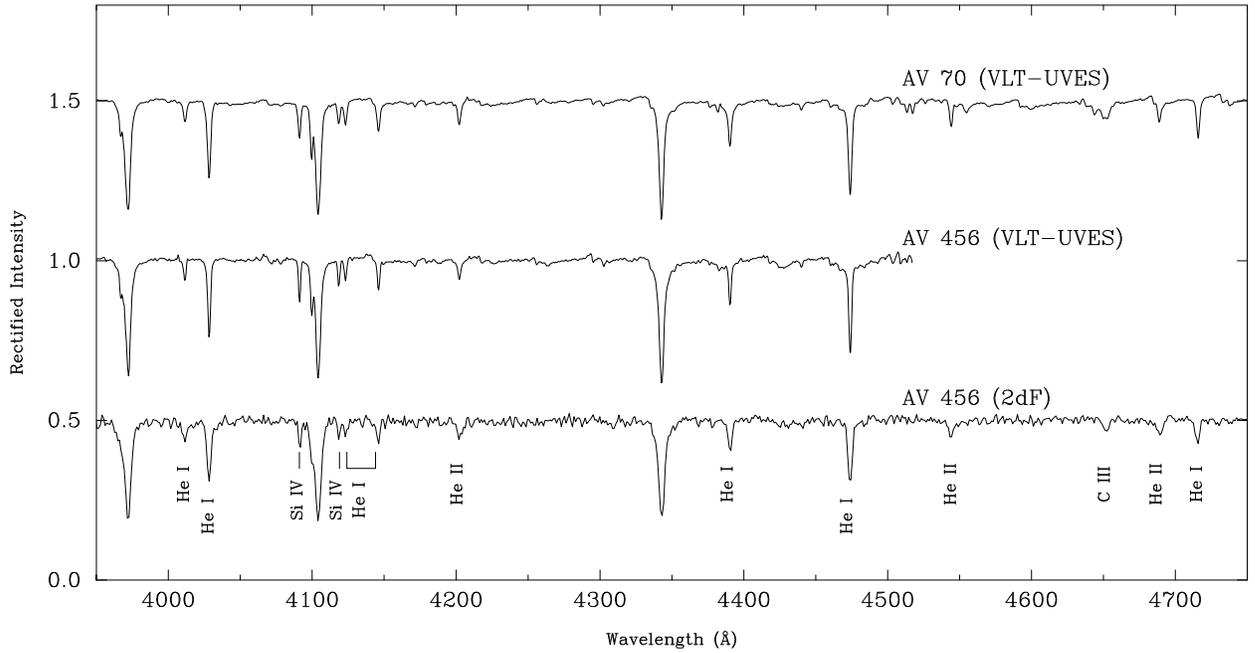}
\caption
{Blue-violet UVES and 2dF spectra of AV\,456 compared with 
the UVES observation of AV\,70.  The features identified in the 2dF spectrum
are, from left to right by ion, He~\1 \lam\lam4009, 4026, 4121, 4144, 4388, 4471, 
4713; He~\2 \lam\lam4200, 4542, 4686; Si~\4 \lam\lam4089, 4116 and C~\3 \lam4650.
For display purposes the UVES data have been smoothed and rebinned to a resolution
of 1\,\AA~FWHM.\label{class}}
\end{center}
\end{figure*}

With the exception of AV\,456, spectral classifications for our
targets are given by Walborn, et al. (1995, 2002).  Previous
classifications for AV\,456 include O9.5 I (Fitzpatrick 1985), B0
(Nandy, Morgan \& Houziaux 1990), O9.5 V (Massey et al. 1995) and O9.5
Ib (Evans et al. 2004).  New digital data for AV\,456 are presented in
Figure \ref{class}.  Unfortunately the wavelength coverage in the UVES
data is somewhat limited so the spectrum is accompanied by an
intermediate-resolution spectrum obtained using the two-degree field
facility (2dF) at the Anglo-Australian telescope (AAT) from Evans et
al. (2004).  To assist classification the UVES data were
Gaussian-smoothed and rebinned to an effective resolution of
1\,\AA~FWHM.  For comparison, the UVES spectrum of AV\,70 is also
shown (classified as O9.5 Ibw by Walborn et al. 2002).  Note the
strong similarity between the two UVES spectra, particularly the ratio
of He~\2 \lam4200 to He~\1 \lam4144 and the intensity of the He~\2 \lam4686
absorption; for these reasons we classify AV\,456 as O9.5 Ibw.  

Morphologically, the ON-type star in our sample, HDE\,269896 (ON9.7\,Ia$^+$) 
is very similar to Sk$-$66$^\circ$169 (O9.7Ia$^+$) from Paper I.  Digital data
for HDE\,269896 were published by Walborn \& Fitzpatrick (1990) and for
Sk$-$66$^\circ$169 by Fitzpatrick (1991).  A useful illustration of
the ON/OC phenomenon in high-quality digital data is given by Walborn
\& Howarth (2000).  The ON classification for HDE\,269896 arises
because of the approximately equal N~\3 \lam4640 and C~\3 \lam4650
intensities, compared to the ``normal'' case in Sk$-$66$^\circ$169, where
the C~\3 absorption is much stronger; the N~\3 \lam4097 line is also
stronger with respect to H$\delta$ in HDE\,269896.  We note that
the peak intensity of the H$\alpha$ emission for HDE\,269896 is larger
than that for Sk$-$66$^\circ$169, suggesting a greater mass-loss rate.

\subsection{Far-UV and UV spectroscopy}
Far-UV spectra for the majority of our targets were obtained from
$FUSE$ Principal Investigator Team programs P117 (P.I.: J. B. Hutchings)
and P103 (P.I.: K. R. Sembach); total exposure times are given in Table 
\ref{targets2}.  Additionally, AV\,70 was observed as part of program
B090 (P.I.: J. M. Shull) and AV\,456 for program Q107 (P.I.: R. Ferlet).
As in Paper I, data from each $FUSE$ detector were processed using
the CALFUSE 2.0.5 pipeline and then subsequently aligned, merged and
resampled to a constant wavelength step of 0.13\,\AA~as described by
Walborn et al. (2002).

To augment the $FUSE$ spectra we relied on UV spectroscopy of
our targets from the $IUE$ and $HST$ archives (as shown in Table
\ref{targets2}).  The majority of our targets were observed with the $IUE$
satellite using the large aperture in the short-wavelength (SWP)
channel at high dispersion (HIRES).  For AV\,469 and 372 we have used
$HST$ Faint Object Spectrograph (FOS) spectra from $HST$ programme
5444 (P.I.: C. Robert), details of which can be found in Leitherer et
al. (2001).

The reddening in the direction of AV\,456 is much larger than that for
the other SMC targets (see Table \ref{targets1}).  Indeed, the $FUSE$
spectrum of AV\,456 contains such significant molecular hydrogen absorption that
it is essentially devoid of useful stellar features apart from the
C~\3 \lam1176 multiplet.  AV\,456 was observed with $IUE$, however
this was at low dispersion (LORES) meaning that our far-UV and UV
coverage for this star is relatively limited in comparison to the rest
of the sample.

\section{Analysis method}
\label{methods}
A comparison between unblanketed, plane-parallel results and
line-blanketed, spherical results for four O-type supergiants was
presented in Paper\,I; further comparisons are not undertaken here.
In contrast, Bouret et al. (2003) demonstrated good agreement between
{\it fully} line-blanketed, plane-parallel models computed with {\sc
tlusty} and {\sc cmfgen} models for O-type dwarfs in the SMC, for
which the effects of the stellar wind on the emergent spectrum are
substantially smaller.

Our methods for the determination of intrinsic parameters were largely
identical to the approach taken in Paper I and by Hillier et
al. (2003).  We varied the mass-loss rate and the velocity law (as
characterized by the exponent $\beta$) until the H$\alpha$ profile was
best reproduced in terms of intensity and morphology, subsequently
adjusting the stellar temperature to match the intensities of the
He~\2 \lam4200/4542 and He~\1 lines. However, He~\2 becomes
exceptionally weak for the B-type supergiants in our sample so the
relative intensities of Si~\4 \lam4089 and Si~\3
\lam4553-68-75 lines were used instead as primary temperature
indicators for these stars.  The alternative temperature diagnostics
were generally in excellent agreement.  

Terminal velocities, $v_\infty$, were primarily determined from
$v_{\rm black}$ (e.g., Prinja, Barlow, \& Howarth, 1990) of the N~\3
\lam990 doublet (largely uncontaminated from molecular hydrogen), 
with consideration to other saturated lines in the far-UV.  Terminal
velocities for each of our targets are listed in Table \ref{results}.

A firm determination of the helium content of extreme supergiants is
extremely difficult (e.g., Hillier et al. 2003) so, following the
arguments from Paper I, the He abundance (by number) was fixed for our
most luminous targets at He/H~$=$~0.2.  For the less luminous
supergiants, we initially assume He/H~$=$~0.2, though we have explored
alternative abundances.  As mentioned in Paper I, helium abundance
differences of $\sim$0.1 are not found to affect the derived
temperatures significantly, although an additional uncertainty is
introduced into the derived mass-loss rate.

CNO abundances were varied to fit the
relevant optical lines, with a typical accuracy of $\sim$0.3 dex.
Solar CNO abundances were taken from Grevesse \& Sauval (1998) except
oxygen, which was set at log(O/H)+12 = 8.66 (Asplund, 2003).
Abundances of other metallic elements were fixed at 0.4$Z_\odot$ (LMC)
and 0.2$Z_\odot$ (SMC), e.g., Russell \& Dopita (1992).  The primary
diagnostic line used for determination of the nitrogen abundance was
N~\3 \lam4097, with consideration of N~\2 \lam3995 and the N~\3
\lam4640 blend which are less sensitive to abundance changes. Oxygen
and carbon abundances were also determined from consideration of the
optical lines; O~\2 \lam4415--17 and \lam4069--4092 provided useful
constraints on the oxygen abundance, whilst C~\2 \lam4267 together
with C~\3 \lam4647--51 (blended with O~\2) were used for carbon.

Another physical parameter that affects the analysis is the rotational
velocity of the star.  This redistributes the flux in a given line,
with the net result that peak absorption and emission intensities are
reduced.  Work is underway by Hillier and co-workers to include the
effects of rotation on stellar wind lines (see discussion in Hillier
et al. 2003; also Busche 2001, Busche \& Hillier 2004). Here we
limited ourselves to the standard procedure of convolving the
synthetic spectrum with a rotational broadening profile.  Following
the method of Herrero et al. (1992), we estimate \vsini~values using
weak metal lines (e.g., those of Si~\3) and the weaker He~\1 lines
(e.g., \lam4009).  The model spectra were convolved initially by
\vsini~=~80 \kms~(rotational velocities in early-type supergiants are
generally low, e.g., Howarth et al. 1997) and, if necessary, then
changed to match the breadth of the metal/He~\1 lines.  Adopted
\vsini~values are given in Table \ref{results}.

The model atoms used for our {\sc cmfgen} calculations were similar to
those used for the Sk$-$66$^{\circ}$169 in Paper I (see Table 4 therein).
Our flux calculations allow for a radially dependent microturbulence\footnote
{In {\sc cmfgen} this adopted value is technically
$\xi_{\rm min}$, the microturbulence at the base of the wind, which
increases with radius to $\xi_{\rm max}$ = 100 \kms~at $v_\infty$.}
($\xi$) as described by Hillier et al. (2003).  In Paper I the most consistent
fit for Sk-66$^\circ$169 (O9.7\,Ia$^+$) was achieved with $\xi$ = 20
\kms; the same value was found by Villamariz et al. (2002) in
their analysis of the Galactic O9.5\,Ib supergiant HD~209975.  For the
present sample we initally assume $\xi$ = 20 \kms, with other values
considered if consistent fits are not found for the helium and silicon lines.

\section{Analyses of Magellanic Cloud OB-type supergiants}
\label{analysis}

Stellar parameters derived from comparisons of the observed spectra
with {\sc cmfgen} models are summarized in Table~\ref{results}.  Three
stars (AV\,235, HDE\,269050 and HDE\,269896) have strong stellar winds
for which unique values of \mdot\, and $\beta$ are determined. In
contrast, the winds of our remaining targets are much weaker and, as
in the case of AV\,69 (Hillier et al. 2003), there can be some
degeneracy between $\beta$ and \mdot\,(as noted by e.g., Puls et al. 1996)
leading to a range of derived parameters for AV\,456 and 469.  The
determination of the best fitting model for each star is subjective,
primarily driven by matching the observed H$\alpha$ profile for each
star.

In most cases, agreement between the H$\alpha$ derived mass-loss rate
and blue visual region is excellent, with the exception of He~\2
\lam4686 (formed both in the photosphere and in the transition zone 
at the base of the wind).  At the temperatures and luminosities of the
current sample \lam4686 demonstrates a wide range of behaviour, i.e.,
from strong absorption (e.g., AV\,70, see Figure \ref{blue_2}) to
strong emission (HDE\,269896).  This line is strongly sensitive to
both the atmospheric extension and the photospheric microturbulence;
larger turbulent velocities drive the line more strongly into
emission.  In comparison with other diagnostic lines, the behaviour of
the model \lam4686 profiles suggests that, for this line, we are
perhaps limited by the adoption of a depth independent photospheric
turbulence.

\begin{figure*}
\begin{center}
\includegraphics[scale=0.7]{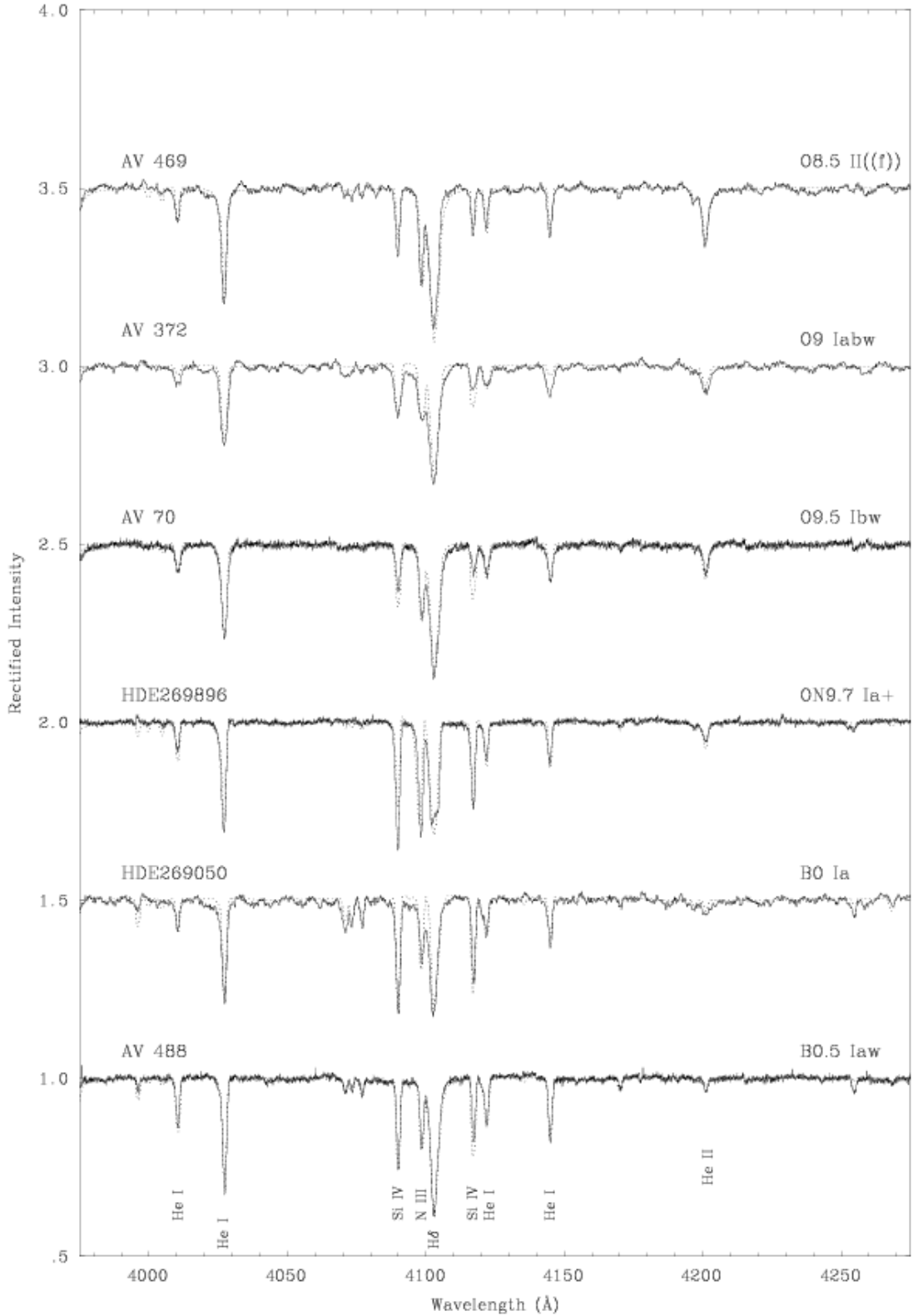}
\caption{\footnotesize \lam4000--4250 region for UVES observations 
(solid line) and the final homogeneous (i.e., unclumped) {\sc cmfgen} models 
(dotted). The lines identified are, from left to right 
by species, He~\1 \lam\lam4009, 4026, 4121,
4144; He~\2 \lam4200; Si~\4 \lam\lam4089, 4116 N~\2 \lam 3995 and N~\3 \lam4097.
To aid the clarity of the comparisons (both here and in the two subsequent figures)
the UVES data and model spectra have been smoothed and rebinned to a resolution of 
1\,\AA~FWHM.  Due to its limited wavelength coverage in the UVES observations, AV\,456
is not included here, nor in Figure \ref{blue_2}.\label{blue_1}}
\end{center}
\end{figure*}

\begin{figure*}
\begin{center}
\includegraphics[scale=0.7]{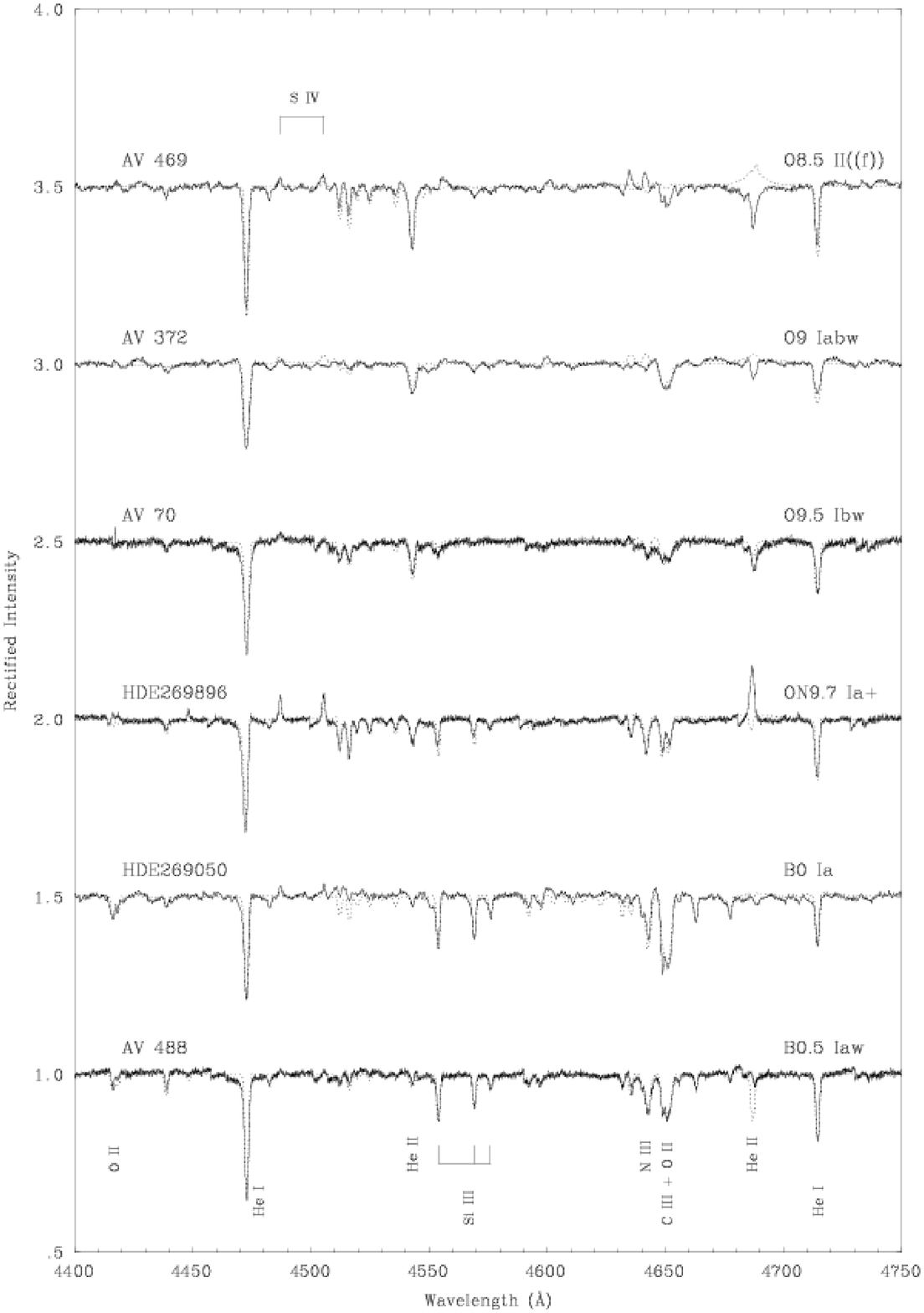}
\caption{\footnotesize \lam4450--4750 region for UVES observations (solid line) 
and homogeneous {\sc cmfgen} models (dotted).  The lines identified in
AV\,488 are, from left to right, O~\2 \lam4415-17, He~\1 \lam4471,
He~\2 \lam4542, Si~\3 \lam4553-68-75, N~\3 \lam4640, C~\3$+$O~\2
\lam4650, He~\2 \lam4686 and He~\1 \lam4713.  The S~\4 \lam\lam4486-4504 
emission lines are also identified in HDE\,269896.\label{blue_2}}
\end{center}
\end{figure*}

\begin{figure*}
\begin{center}
\includegraphics[scale=0.7]{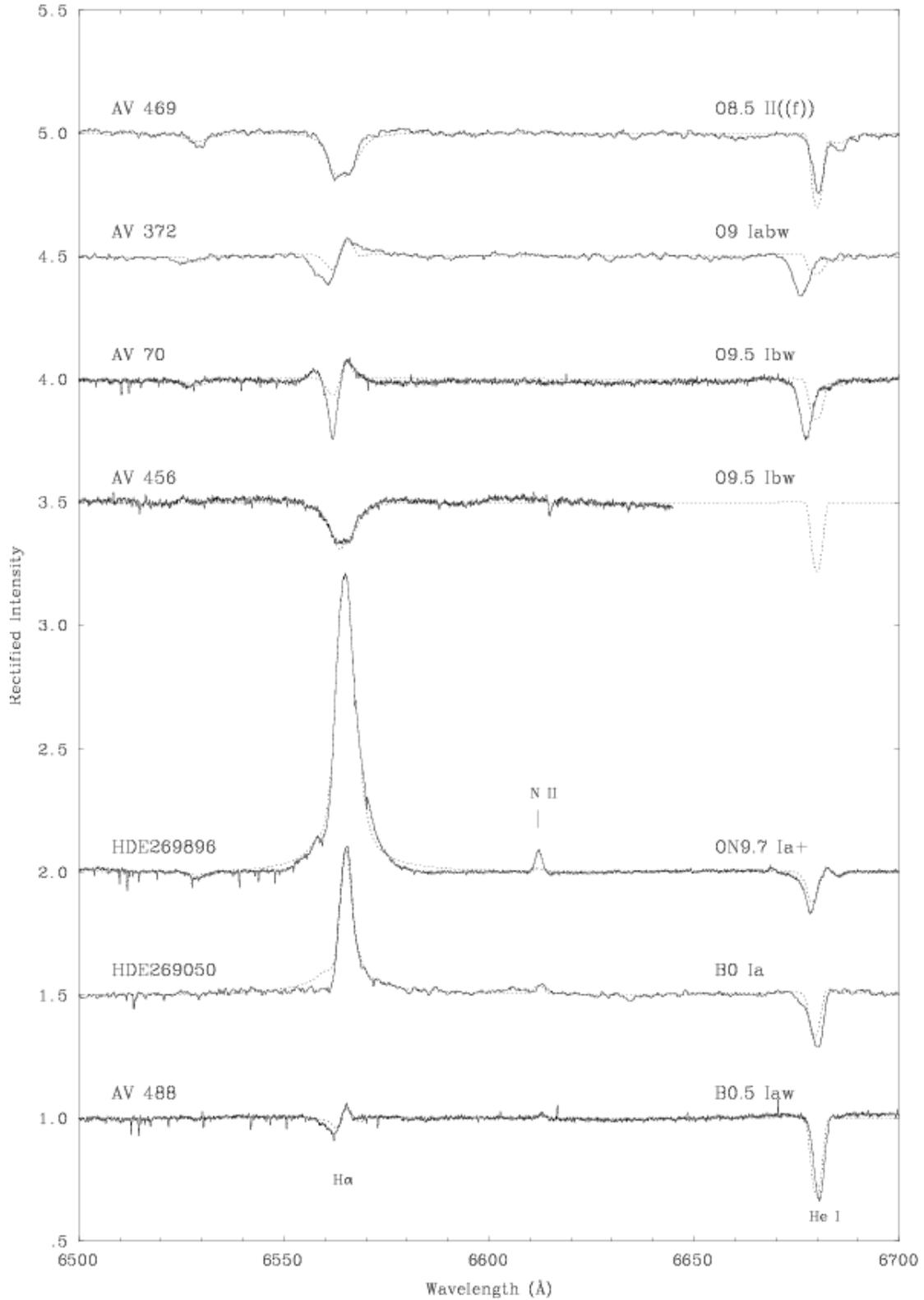}
\caption{\footnotesize \lam6500--6700 region for UVES observations (solid lines) and
homogeneous {\sc cmfgen} models (dotted).  The He~\1 \lam6678 line is
present in each spectrum and N~\2 \lam6611 emission is seen in both
HDE\,269896 and 269050.  The velocity offsets between the H$\alpha$ and He~\1 
line are discussed in the text.\label{halpha_3}}
\end{center}
\end{figure*}

UV and far-UV comparisons are generally successful, as we shall
discuss later. However, for AV\,235 the agreement between H$\alpha$
and the other Balmer lines is extremely poor; it is discussed
separately in \sref{av235}.  The remaining stars are now discussed in turn,
with synthetic spectra compared to optical observations in
Figures~\ref{blue_1}--\ref{halpha_3}. One further unresolved problem
is common to most stars.  As revealed in Figure~\ref{halpha_3}, there
is often a substantial $\sim$50--100 \kms~offset between the apparent
radial velocity shift of the observed H$\alpha$ profile compared to
that of the He~\1 \lam6678 line. The radial velocity of the He~\1 line
is more in keeping with the blue data (see Table \ref{targets1}),
i.e., the peak of the H$\alpha$ emission is significantly discrepant
from its expected wavelength in the $observed$ spectrum.  Checks of
the UVES observations with conventional long-slit spectra taken with
the 2.3m Australian National University (ANU) telescope, confirm the
wavelength of the observed H$\alpha$ feature in all cases, and
measurements of the He~\1 \lam7065 line are consistent with those at
\lam6678.  Although the effect was less significant ($\sim$40 kms),
a similar offset was also present in the spectrum of AV\,83 (Hillier et
al. 2003).

\subsection{AV\,469 [Sk~148, O8.5 II((f))]}

We derive \Teff~= 33\,kK, $\log (L/L_\odot)$ = 5.50 and $v_\infty$ = 1550 
\kms\,for AV\,469, very similar parameters to those found for AV\,69 
[OC7.5\,III((f))] by Hillier et al. (2003).  The H$\alpha$ profile of
AV\,469 (see Figure \ref{halpha_3}) is strongly in absorption and there is
some degeneracy between \mdot~and $\beta$.  Models were calculated for
$\beta$ exponents in the range 0.85--1.75, varying the mass-loss rate
to optimally match the H$\alpha$ profile.  The best fitting model has
\mdot~=1.3$\times$10$^{-6}$  $M_\odot$ yr$^{-1}$
with $\beta$~=~1.0, although reasonable fits
can also be obtained adopting $\beta$ = 0.85 and 1.25, resulting in
\mdot~values of order 20$\%$ higher and lower respectively.  The other
derived parameters remain invariant to such changes in $\beta$.

As with our other targets these models assume He/H = 0.2; lower He
abundance models were generated but provided less consistent fits to
the helium lines.  The nitrogen abundance is fixed using N~\3
\lam4097, revealing a mass fraction (relative to solar) of 
$\epsilon$(N/N$_\odot$)$\sim$1.5, indicating a large enhancement when
compared to the SMC interstellar medium.  We are unable to match the
observed 4634--41 N~\3 emission lines, although in our models they are
somewhat ``filled-in'' (cf. the absorption in our models for other
stars, see Figure \ref{blue_2}).  The carbon abundance derived using
C~\3
\lam4650 for AV\,469 is $\epsilon$(C/C$_\odot$) = 0.03, providing
strong evidence of depletion compared to the nebular abundances.  As
one might expect given the relatively high stellar temperature, the
O~\2 \lam4415-17 doublet is not detected in AV\,469, so no firm
abundance determination was possible.

\subsection{AV\,372 (Sk 116, O9 Iabw)}

From inspection of the H$\alpha$ profile (see Figure \ref{halpha_3})
AV\,372 clearly has a very weak wind. The final model has \Teff =
28\,kK, $\log(L/L_\odot) = 5.62$ and \mdot = 1.0$\times$10$^{-6}$ 
 $M_\odot$ yr$^{-1}$ with
$\beta$ = 2.25 and a helium number abundance of He/H = 0.15.  The
terminal velocity from the $FUSE$ spectrum is $v_{\infty} = 1550$ \kms,
identical to that found from $HST$ UV spectra by Prinja \& Crowther
(1998).  We are unable to simultaneously match the observed absorption
and weak emission; lower values of $\beta$ with a higher mass-loss
rate (and vice versa) do not successfully improve agreement with the
observed profile.  As mentioned earlier, there is a velocity offset
between H$\alpha$ and the He~\1 lines in the red UVES data, in this
case $\sim$100 \kms.

Again, nitrogen is strongly enhanced relative to the ISM of the SMC, with
$\epsilon$(N/N$_\odot$) = 0.6.  The O~\2 \lam4415-17 doublet is not
visible in the AV\,372 UVES spectrum, thus the \lam4650 feature is
expected to originate largely from C~\3, leading to the carbon
abundance of $\epsilon$(C/C$_\odot$) = 0.05.

\subsection{AV\,70 (Sk~35, O9.5 Ibw)}
\label{av070}

The H$\alpha$ profile in AV\,70 (see Figure \ref{halpha_3}) has a
double emission peak with significant central absorption, a feature
commonly seen in Be-type stars (e.g., Slettebak, 1988).  We are unable
to reproduce this feature using our current 1D method and limit
ourselves here to simultaneously matching the redward emission and the
blue-region optical lines.  The parameters found for AV\,70 are
\Teff = 28.5\,kK, $\log (L/L_\odot)$ = 5.68, 
\mdot = 1.5$\times$10$^{-6}$ $M_\odot$ yr$^{-1}$ with $\beta$ = 1.75.
Again note the observed velocity offset in Figure \ref{halpha_3}
between H$\alpha$ and He~\1 \lam6678 of $\sim$100\kms.

As with AV\,469 and AV\,372, nitrogen is over-abundant in AV\,70, with
$\epsilon$(N/N$_\odot$) $\sim$ 0.9, whilst carbon is deficient,
$\epsilon$(C/C$_\odot$) = 0.03.  Given the relatively high stellar
temperature for the preferred AV\,70 model, the O~\2 \lam4415-17 is
very weak (if present at all) and the derived oxygen abundance is
given as an upper limit.  

We are unable to match all of the observed helium and silicon features
simultaneously for AV\,70; the Si~\4 absorption lines in the final
model are too strong (see Figure \ref{blue_1}).  The adopted
microturbulence (i.e., $\xi = 20$ \kms) was that found from
consideration of the helium lines by, e.g., Villamariz et al. (2002).
In early B-type stars the microturblent velocity is generally
determined from silicon (and oxygen) lines and the result values
are typically lower ($\sim$10 \kms, e.g., McErlean et al. 1999).
Calculation of the formal solution for AV\,70 with $\xi = 10$
\kms\,gives good agreement for the Si~\3/\4 and He~\2 lines, with the
consequence that the predicted intensity of the He~\1 lines is
generally too small.  This highlights the potentially different values
of $\xi$ obtained from separate elements, as noted by McErlean et
al. and Vrancken et al. (2000) in reference to results from silicon
and oxygen.

\subsection{AV\,456 (Sk 143, O9.5 Ibw)}
\label{av456_2}

As discussed in \sref{obsdetails}, the blue optical spectrum of
AV\,456 is strikingly similar to that of AV\,70 and one might
therefore anticipate comparable stellar parameters.  However, the
H$\alpha$ profiles of the two stars differ substantially (see Figure
\ref{halpha_3}).  Due to the high interstellar extinction, we were
unable to determine $v_\infty$ for AV\,456 using the N~\3 and other
far-UV lines, so we adopt the same terminal velocity as for AV\,70
(recall only LORES $IUE$ datasets are available for this star). 

An optimal fit was obtained for \Teff = 29.5\,kK and \mdot~=
0.7$\times$10$^{-6}$ $M_\odot$ yr$^{-1}$ with $\beta$ = 1.75.
The blue visual data for AV\,456 are omitted from Figures~\ref{blue_1}
and \ref{blue_2} as they are much noisier than those for the
other targets.

A helium number abundance of He/H = 0.1 is preferred for AV\,456,
although in this instance such differences in helium abundance
(cf. AV\,70) should not be considered significant.  Recall that the
H$\alpha$ profile for AV\,70 is poorly reproduced in our spherical 1D
model -- it is possible that a circumstellar disk or shell leads to
the observed emission superimposed on the stellar absorption
profile. This would lead to an artificially high mass-loss rate and
therefore the higher helium abundance.  From inspection of the raw
data the He~\1 lines in AV\,70 are slightly broader and shallower than
in AV\,456, accounting for the slight difference in the adopted
\vsini~values.

Due to the lack of optical data in the \lam4500-4700\,\AA~region, a
thorough abundance analysis is not possible for AV\,456, although
$\epsilon$(N/N$_\odot$) = 0.6 is derived for nitrogen from N~\3
\lam4097.  The O~\2 \lam4415-17 doublet (if present) is
indistinguishable from the noise and so the upper limit to the oxygen
abundance is again $\epsilon$(O/O$_\odot$) $\leq$ 0.2.

\subsection{HDE\,269896 (Sk--68$^{\circ}$135, ON9.7 Ia$^+$)}

The terminal velocity found from fits to the $FUSE$ lines ($v_\infty =
1350$ \kms) for HDE\,269896 is somewhat larger than that found by
Massa et al. (2003) from analysis of $IUE$ spectra ($v_\infty = 1050$
\kms).  For consistency with the methods employed for the rest of the current
sample (and those in Paper I) we adopt the value found from the $FUSE$
data.

HDE\,269896 is significantly more luminous than its spectroscopic twin
(Sk$-$66$^\circ$169) with $\log (L/L_\odot)$ = 5.97.  We derive a
temperature of \Teff~=~27.5\,kK, with a mass-loss rate of \mdot =
7.5$\times$10$^{-6}$ $M_\odot$/yr$^{-1}$ and a relatively large wind
exponent of $\beta = 3.5$ (i.e., a comparatively slow radial
acceleration).  We find evidence of nitrogen enhancement with
$\epsilon$(N/N$_\odot$) $\sim$ 2, whilst both carbon and oxygen are
metal poor with $\epsilon$(C/C$_\odot$) $\sim$ 0.06 and
$\epsilon$(O/O$_\odot$) $\sim$ 0.17. A comparison between HDE\,269896
and Sk$-$66$^\circ$169 is made in Sect.~\ref{abundances}.

\subsection{HDE\,269050 (Sk--68$^{\circ}$52, B0 Ia)}

The derived parameters for HDE\,269050 are \Teff = 24.5\,kK, $\log
(L/L_{\odot}) = 5.76$, \mdot = 3.2$\times$10$^{-6}$ 
 $M_\odot$ yr$^{-1}$with $\beta$ =
2.75.  The discrepancy between the radial velocities in the observed
H$\alpha$ profile and He~\1 \lam6678 is also present in HDE\,269050,
though its magnitude is lower ($\sim$60 \kms, see Figure \ref{halpha_3}).  Again there is
evidence of significant CNO-processing with $\epsilon$(C/C$_\odot$) = 0.15
$\epsilon$(N/N$_\odot$) = 3.7 and $\epsilon$(O/O$_\odot$) = 0.5,
the latter based on optical O~\2 diagnostics.

\subsection{AV\,488 (Sk~159, B0.5 Iaw)}
The redward emission in the H$\alpha$ profile of AV\,488 permits a
unique determination of \mdot~and $\beta$.  The final derived
parameters for AV\,488 are \Teff = 27\,kK, $\log (L/L_\odot)$ = 5.74,
\mdot~= 1.2$\times$10$^{-6}$ $M_\odot$ yr$^{-1}$ with $\beta$ = 1.75,
and He/H = 0.20.  The terminal velocity from the $FUSE$ data is
$v_\infty = 1250$~\kms, compared to previous determinations of 1300
\kms~(Haser 1995) and 1040 \kms~(Prinja \& Crowther 1998).  There is 
also a velocity shift between He~\1 \lam6678 and H$\alpha$ of $\sim$60
\kms (see Figure \ref{halpha_3}), though in the opposite sense to that seen in our other targets, 
i.e., the observed H$\alpha$ profile is apparently consistent with a 
lower radial velocity than that from He~\1.  Models were also
calculated with He/H = 0.1 by number; the derived parameters of the best
fitting model were identical, with the exception that \mdot~=
1.0$\times$10$^{-6}$ $M_\odot$ yr$^{-1}$ and the quality of the fits
to the He~\1 lines was slightly diminished. 
Once again, based on several N~\2, C~\2 and O~\2 lines, AV\,488 shows evidence
of  partial CNO-processing,  with $\epsilon$(C/C$_\odot$)
= 0.04, $\epsilon$(N/N$_\odot$) = 1.2 and $\epsilon$(O/O$_\odot$) =
0.17.

\subsection{UV and far-UV comparisons: 
Further evidence for clumping in OB-type stars?}

Thus far, we have carried out a comparison between synthetic {\it optical} 
spectra and observations. As in Paper~I, we now consider how
well these optically derived parameters reproduce the UV and far-UV
spectral features. Successes and failures of the present models are
generally common to stars of each spectral type (recall that AV\,235
is discussed separately).  The N\,\5 $\lambda\lambda$1238--42 and O~\6
\lam\lam1032--1038 doublets are purposefully omitted from these
comparisons since these are super-ions produced by X-rays in late O
and early B-type stars (see Paper~I).  Furthermore, HDE\,269896 is
closely reminiscent of Sk$-$66$^\circ$169 from Paper~I and is 
not discussed here.

\begin{figure*}
\begin{center}
\includegraphics[scale=0.55, angle=270]{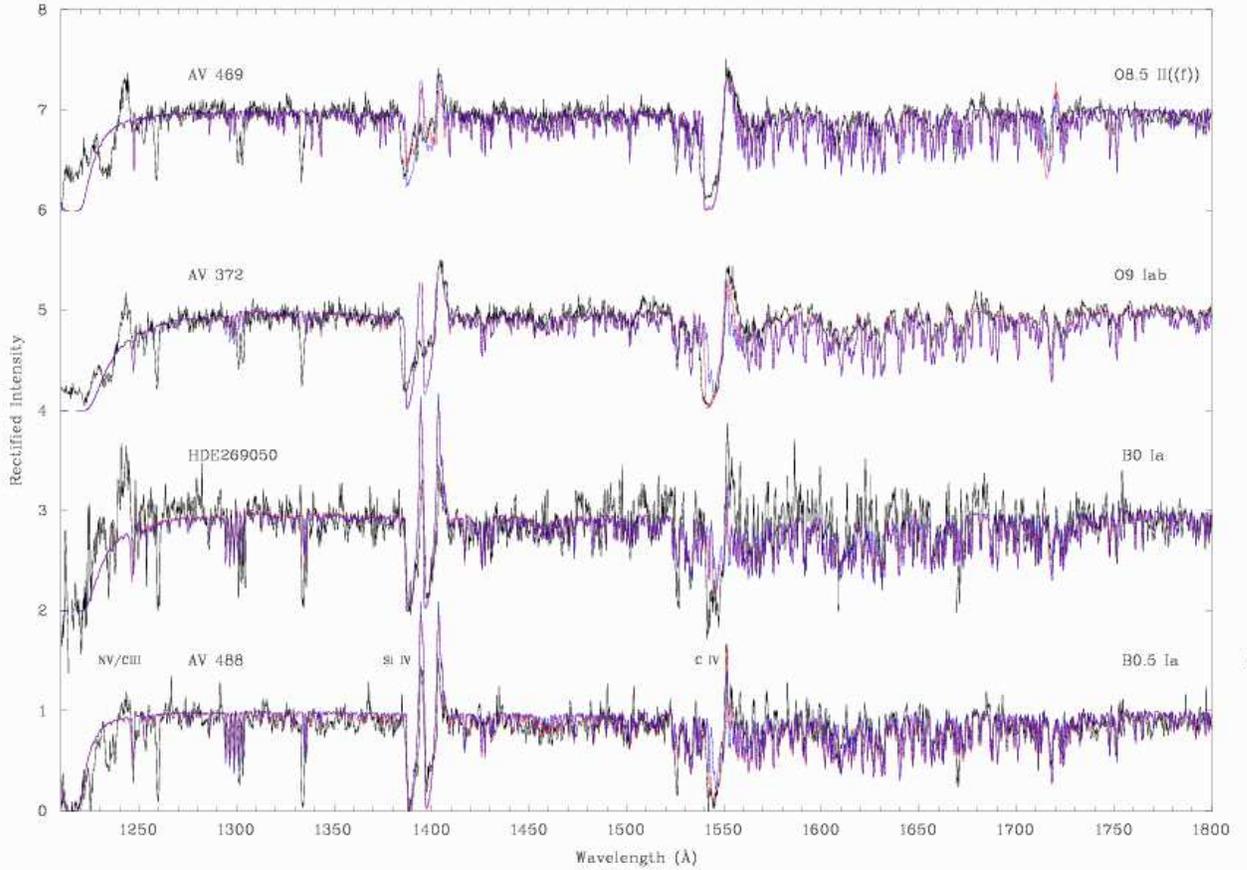}
\caption{Illustrative comparison of $HST$-FOS (AV\,469 \& 372) and $IUE$ (HDE\,269050 \& AV\,488)
data with the final {\sc cmfgen} models (unclumped -- red; clumped -- blue).  
The two lines identified in AV\,488 are Si~\4 \lam\lam1394, 1403 and C~\4 \lam\lam1548-51. 
The model spectra were multiplied by synthetic transmission spectra to incorporate the effects
of absorption by interstellar neutral hydrogen, as measured from Lyman-$\beta$ (Table \ref{targets1}).\label{iue}}
\end{center}
\end{figure*}

In Figure~\ref{iue} we compare model spectra of four representative
targets with $IUE$ observations, covering $\lambda\lambda$1250--1800.
In general, the UV comparison supports the stellar parameters derived
from our optical analysis, although recall from Paper~I that UV wind
diagnostics provide relatively poor discriminants between different
models for O-type spectra (although the photospheric lines are
more useful, e.g., the Fe~\4/Fe~\5 ratio).  The principal spectral features in the
$IUE$ range are Si\,\4 $\lambda\lambda$1393--1402 and C\,\4
$\lambda\lambda$1548--51, together with the iron ``forest'' redward of
$\lambda$1500.  The optically derived models match the observations
reasonably well, with the exception of the Si\,\4 P-Cygni emission,
over-predicted in many of our targets.  The UV plays a greater diagnostic 
role  for early B supergiants, in which Massa (1989) found that
some of the UV silicon lines are sensitive to temperature and
luminosity.  For the later subtypes in our sample (i.e., HDE\,269050 and
AV\,488) there is good agreement in the Si~\3 lines discussed by Massa
(namely \lam1294, 1299 and 1417), which offers further confidence 
in our derived temperatures.

\begin{figure*}
\begin{center}
\includegraphics[scale=0.55, angle=-90]{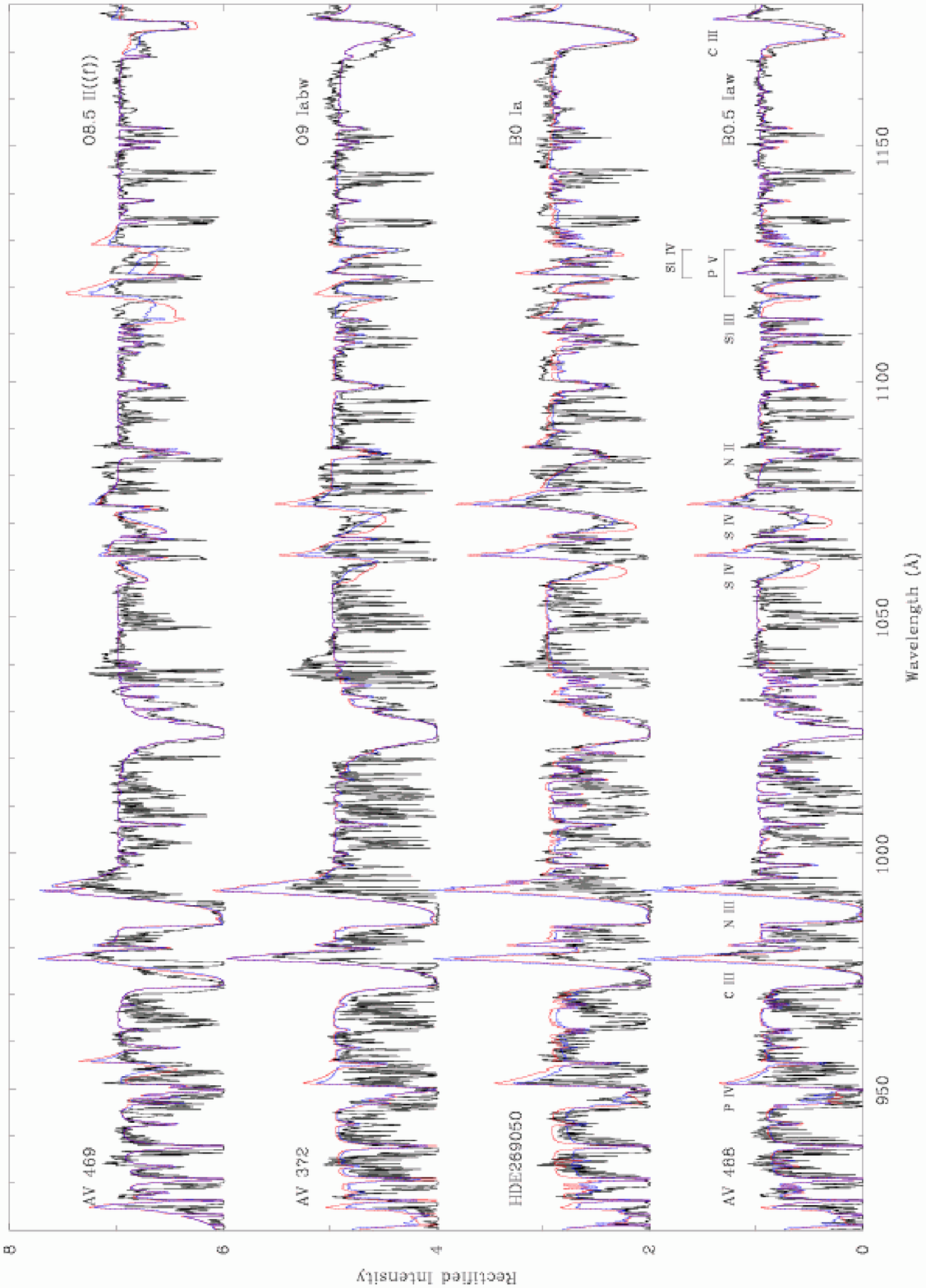}
\caption{Comparison of $FUSE$ data and final {\sc cmfgen} models (unclumped -- red; 
clumped -- blue).  The lines identified in AV\,488 are P~\4 \lam951, C~\3 \lam977, N~\3 \lam991, 
S~\4 \lam\lam1063-73, N~\2 \lam1085, Si~\3\lam\lam1108-10-13, P~\5 \lam\lam1118-28 
and Si~\4 \lam\lam1123-28.  As in Figure \ref{iue}, the model spectra have been multiplied by
an appropriate transmission spectrum to include the effects of absorption by interstellar neutral hydrogen.\label{fuse}}
\end{center}
\end{figure*}

Now we turn to the far-UV $FUSE$ region, which proved much more useful
for the O-type stars in Paper~I.  Unfortunately, interstellar molecular H$_2$ is rather
more problematic for AV\,488 and HDE\,269050 with column densities of
10$^{19-19.5}$ cm$^{-2}$ (Tumlinson et al. 2002) but the 
principal spectral features remain relatively clear (see Figure~\ref{fuse}). 
Many of the far-UV features (e.g., C\,\3~\lam1176 and the O~\3
\lam1139-50-51-54 lines) are well reproduced
by our (unclumped) models, however the P-Cygni emission is again
over-predicted in some lines, e.g., C\,\3 \lam977, N\,\3 \lam991 and the
S\,\4 $\lambda$1062--1072 doublet.

Using the filling factor approach described in Hillier et al. (2003)
we calculated clumped models to investigate the effects on our
diagnostic lines.  The volume filling factor was reduced from 100$\%$
to 10$\%$ (i.e., $f_\infty$ = 0.1) with the velocity at which the
clumping is initiated ($v_{cl}$) set to 30 \kms~(taken from the
analysis of AV\,83 by Hillier et al.).  Matching the H$\alpha$
profiles of our targets with clumped models typically necessitated
mass-loss rates a factor of 3 lower, with all other input parameters
largely unchanged.

\begin{figure*}
\begin{center}
\includegraphics[scale=0.55, angle=-90]{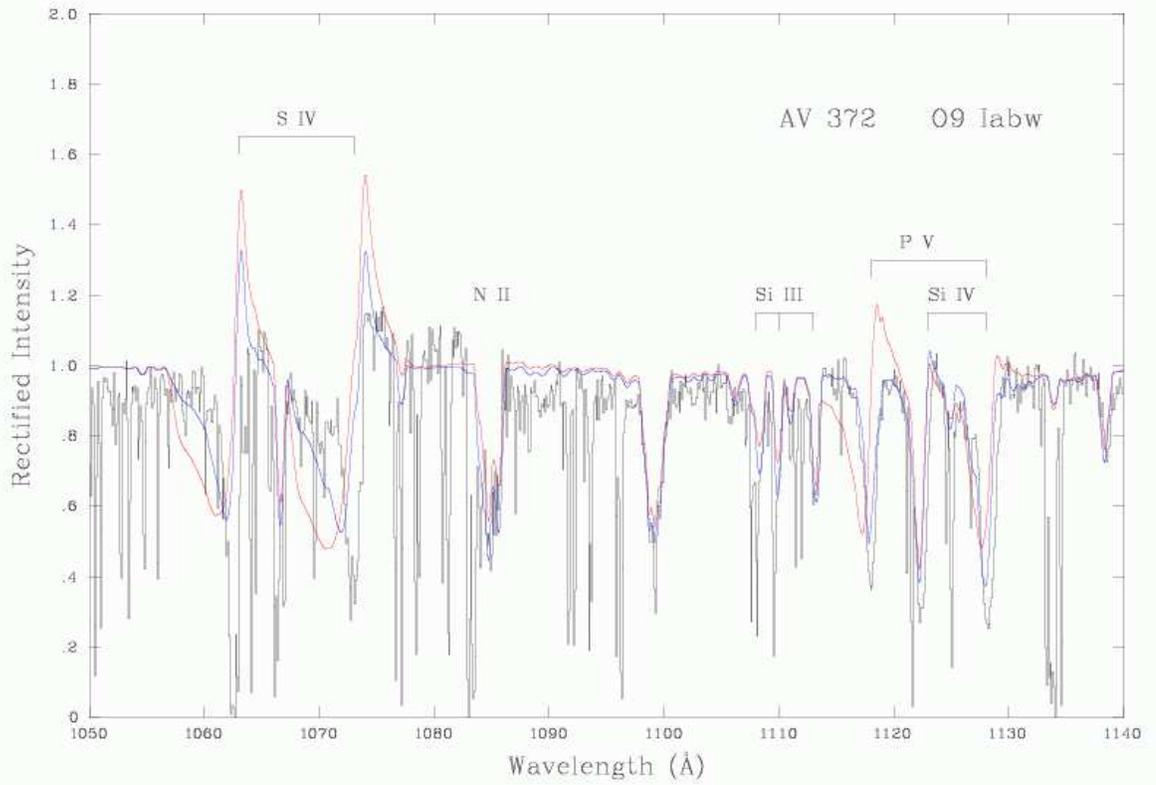}
\caption{Comparison of the observed S~\4 \lam\lam1063-73 and P~\5 \lam\lam1118-28 doublets 
with the final {\sc cmfgen} models (unclumped -- red; clumped -- blue) for
AV\,372.  Additional identified lines are N~\2 \lam1085, Si~\3
\lam\lam1108-10-13 and Si~\4 \lam1123-28.\label{fuse2}}
\end{center}
\end{figure*}

Clumped models are also shown in Figure~\ref{fuse}.  Recall from
Paper~I that P~\5 $\lambda$1118--28 was identified as a potential
indicator of clumping in early-mid O-type stars, both AV\,469 and 372
appear to again demonstrate this.  We suggest here that for late O and
early B-type stars S~\4 $\lambda\lambda$1062--1072 may offer a similar
diagnostic; these lines will be sensitive to clumping as they are
unsaturated and arise from the dominant ion of the species.  In
Figure~\ref{fuse2} we show the S~\4 and P~\5 region for AV\,372 in
more detail.  In contrast to phosphorus, the present day ISM abundance
of sulphur in the Magellanic Clouds is well known from both H\,\2
regions (e.g., Russell \& Dopita 1992) and stars (e.g., Rolleston et
al. 2003).  With the sulphur abundance fixed, a moderately clumped
wind leads to less discrepant emission P-Cygni emission and the
morphology of the absorption components are better matched.  

The role of clumping in the winds of Wolf-Rayet stars is well
documented (e.g., Moffat et al. 1988; Hillier 1991); only more
recently have such effects been considered in O and B-type stars.  We
conclude that it is likely that the winds of O and early B-type
supergiants are clumped, with the best spectroscopic evidence revealed
by unsaturated lines of dominant ions in the far-UV $FUSE$ region.
Massa et al. (2003) arrived at a similar conclusion based on their
line profile analyses of $FUSE$ observations of O-type stars in the
LMC.  Additional supporting evidence was offered by Hillier et al.
(2003) who noted that unclumped models predict asymmetries in some UV
photospheric lines that are not seen in the observations; these
discrepancies are reduced by the inclusion of clumping.  Intensive
spectroscopic monitoring of both optical emission features (Eversberg
et al. 1998) and UV lines (Prinja, Massa, \& Fullerton, 2002) have
already indicated that winds in early type stars are highly
structured; hydrodynamic models (Owocki et al. 1988) and X-ray
spectroscopy (Miller et al. 2002) also suggest wind clumping.

\section{The peculiar case of AV\,235 (Sk~82, B0\,Iaw)}
\label{av235}

\begin{figure*}
\begin{center}
\includegraphics[scale=0.55, angle=-90]{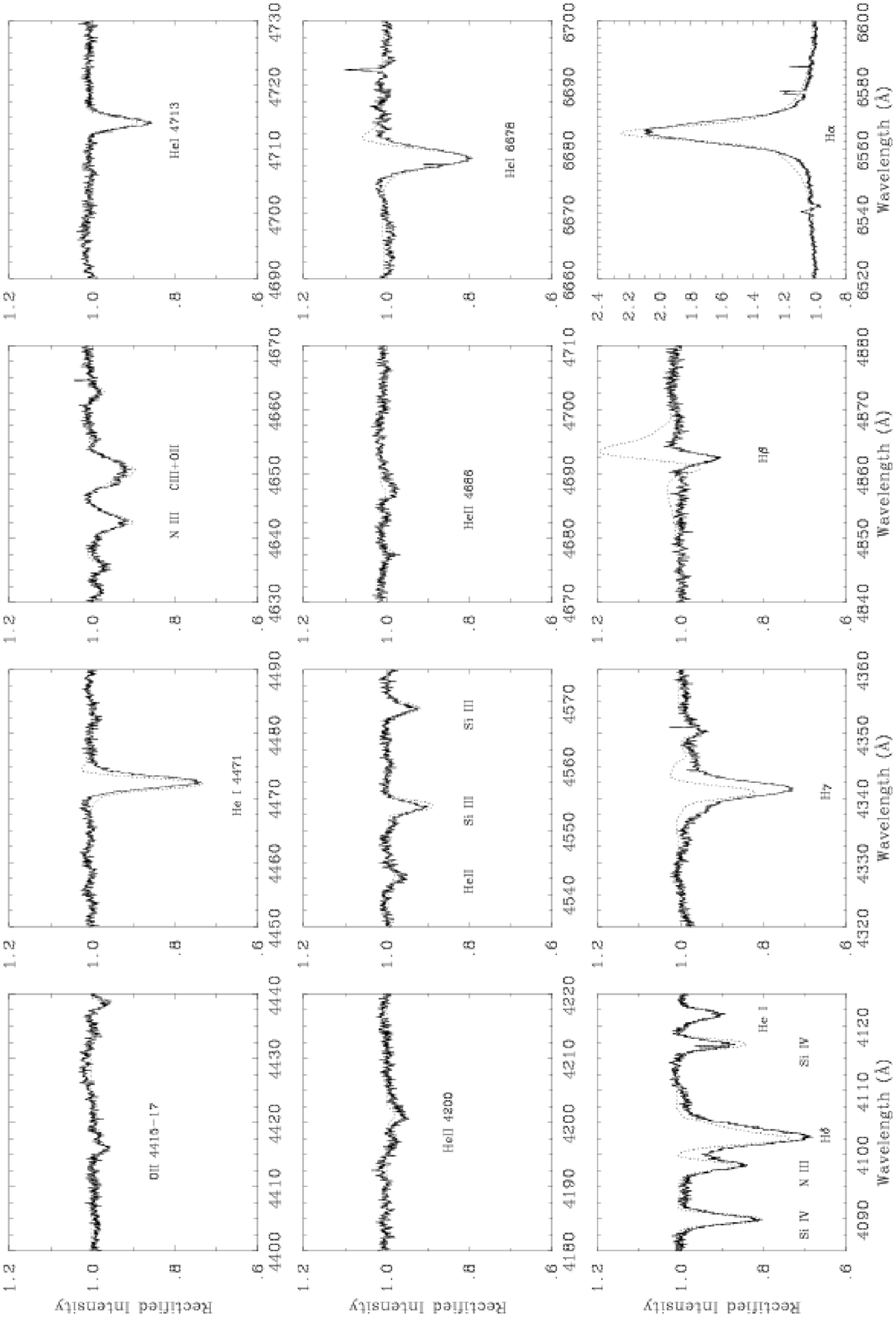}
\caption{Comparison between optical UVES line profiles of AV\,235 (solid line) and 
{\sc cmfgen} spectra (unclumped -- dotted; clumped -- dashed) for our
H$\alpha$ derived mass-loss rate.\label{av235_opt}}
\end{center}
\end{figure*}

For seven of the eight program stars, we obtain reasonably good
agreement between the synthetic far-UV, UV and optical spectra and
observations; for AV\,235 this is not the case.  Following our usual
method, i.e., selecting a mass-loss rate from H$\alpha$ and
temperature from photospheric helium or silicon diagnostics we derive
the following parameters: \Teff~$=$~24.5kK, $\log (L/L_{\odot}) =
5.63$, $\dot{M} = 5.8 \times 10^{-6} M_{\odot}$yr$^{-1}$ with a
$\beta=2.5$ velocity exponent. The synthetic spectrum from this model
is compared to observations in Figure \ref{av235_opt} (to permit a
clearer comparison of the predicted profiles the 100~\kms~offset
between He~\1 \lam6678 and H$\alpha$ has been removed). For this set
of parameters, we obtained CNO abundances of $\epsilon$(N/N$_\odot$) =
2.3, $\epsilon$(O/O$_\odot$) = 0.2 and $\epsilon$(C/C$_\odot$) = 0.02.
The helium lines are well fit by this model, as are the Si~\3 and \4
lines.  However, it is clear that the predicted P Cygni emission
profiles for H$\beta$, H$\gamma$ are far too strong in emission, in
spite of H$\alpha$ matching well; clumping is not able to resolve
these differences.  

\begin{figure*}
\begin{center}
\includegraphics[scale=0.55, angle=-90]{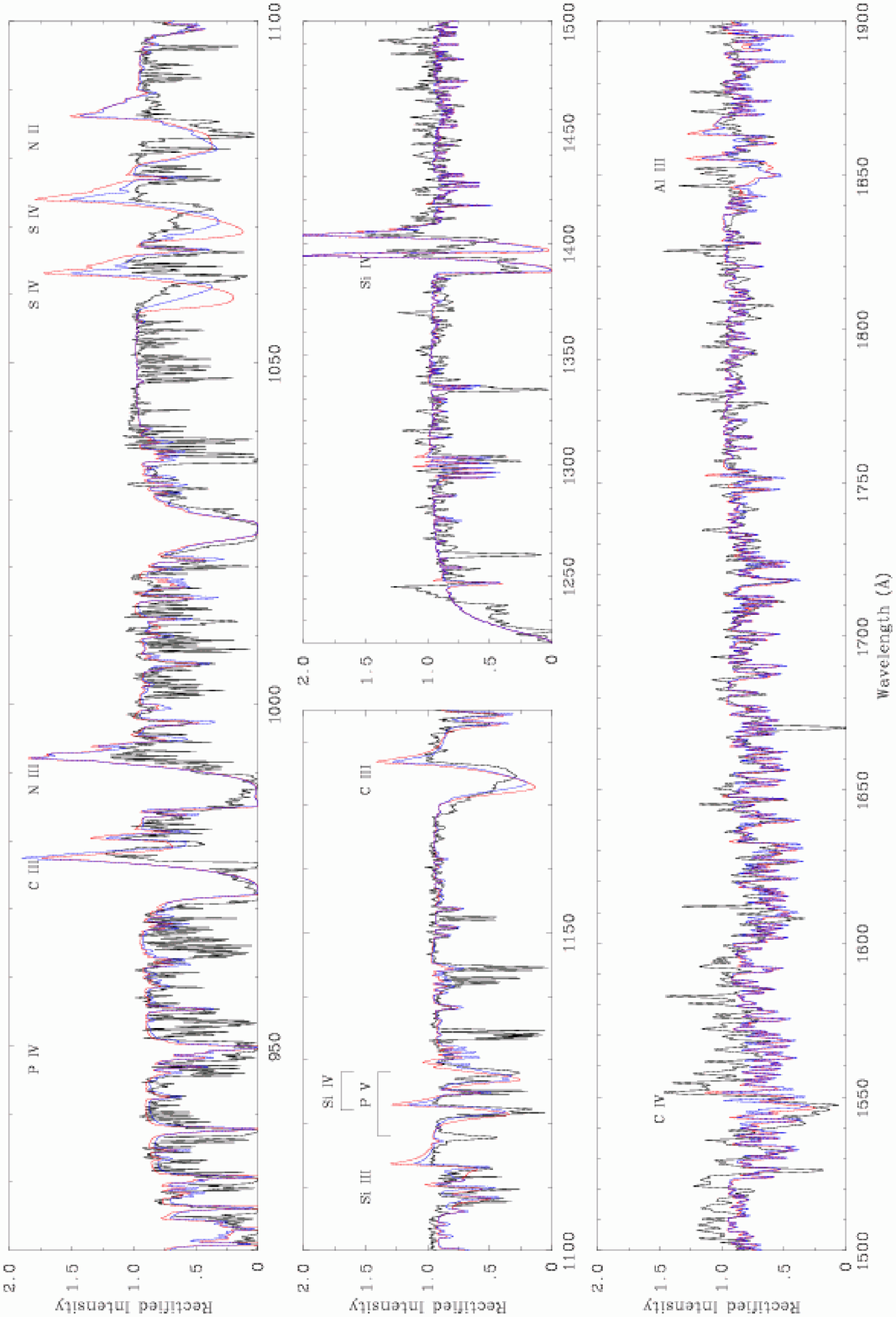}
\caption{Comparison between $FUSE$/$IUE$ observations of AV\,235 and 
the H$\alpha$ derived {\sc cmfgen} model spectra (unclumped -- red; clumped -- blue).
The model spectra have been multiplied by an appropriate transmission spectrum to include
the effects of neutral hydrogen absorption.\label{av235_uv}}
\end{center}
\end{figure*}

Turning to the UV, many of the wind features are
again over-estimated (see Figure \ref{av235_uv}); they too suggest a
much lower mass-loss rate and/or higher ionization than that obtained
from fits to H$\alpha$. For example, N~\2 and Al~\3 are predicted
strongly in emission (yet such features are not observed in B
supergiants earlier than B0.7, e.g., Walborn, Parker, \& Nichols,
1995) whilst the predicted C~\4 is too weak.

\begin{figure*}
\begin{center}
\includegraphics[scale=0.55, angle=-90]{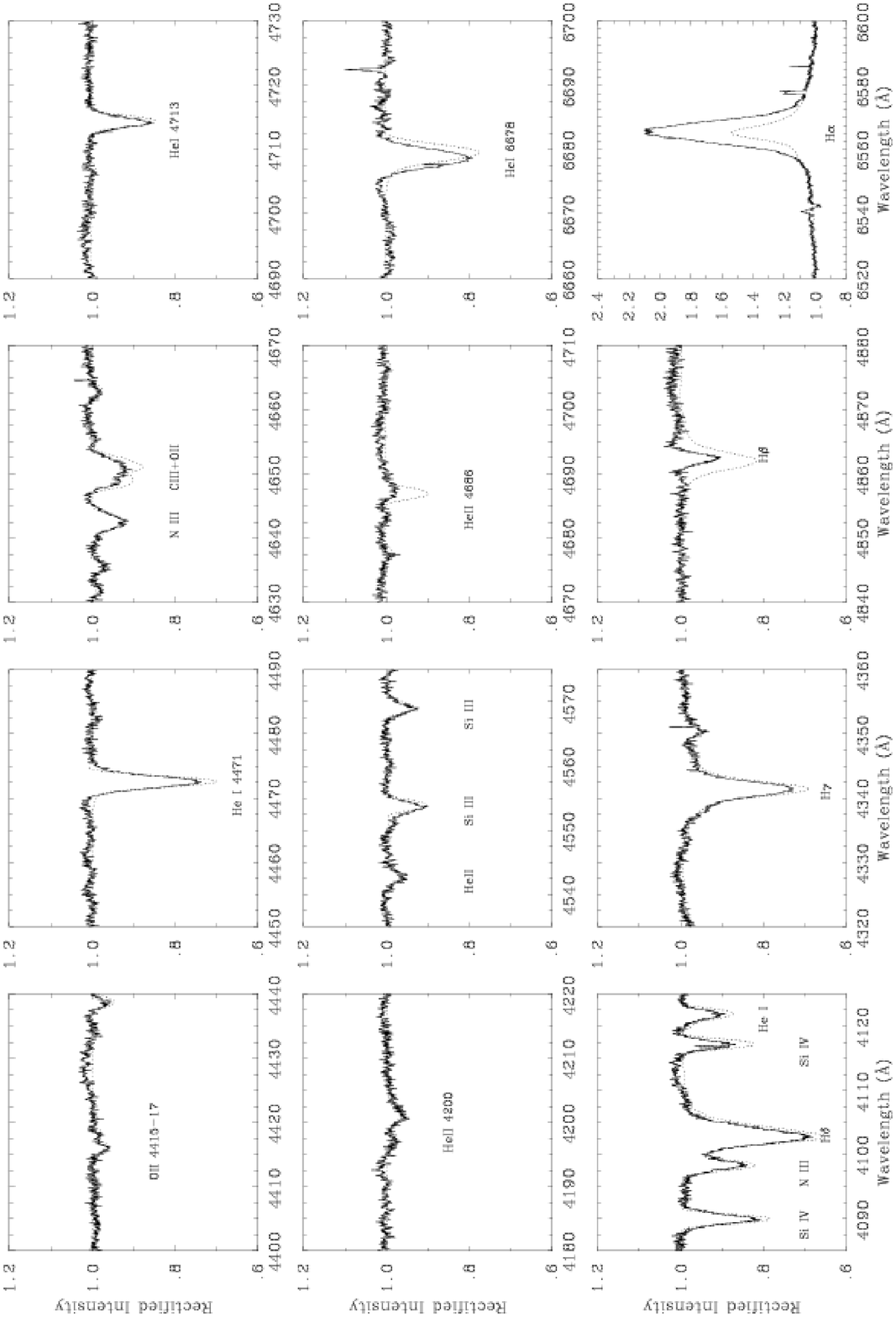}
\caption{Comparison between optical UVES line profiles of AV\,235 (solid line) and {\sc cmfgen} 
spectra (unclumped -- dotted; clumped -- dashed) for our H$\gamma$ derived mass-loss rate.\label{av235_opt2}}
\end{center}
\end{figure*}

We therefore calculated models which reproduced the usual optical
diagnostics, except that H$\gamma$ was now selected as the mass-loss
diagnostic.  In this case (shown in Figure~\ref{av235_opt2}) a
significantly higher stellar temperature was required to reproduce the
photospheric He and Si lines, i.e., \Teff~$=$~27.5kK, $\log
(L/L_{\odot}) = 5.72$, $\dot{M} = 4.0 \times 10^{-6}
M_{\odot}$yr$^{-1}$ with an adopted $\beta=1.5$ (``AV\,235 H$\gamma$''
in Table \ref{results}).  The line wings of H$\gamma$ are now well
matched (though the predicted core absorption is too large) and the
overall optical agreement is equal to the H$\alpha$ model with the
exception that this line and H$\beta$ are now severely
underestimated.  

\begin{figure*}
\begin{center}
\includegraphics[scale=0.55, angle=-90]{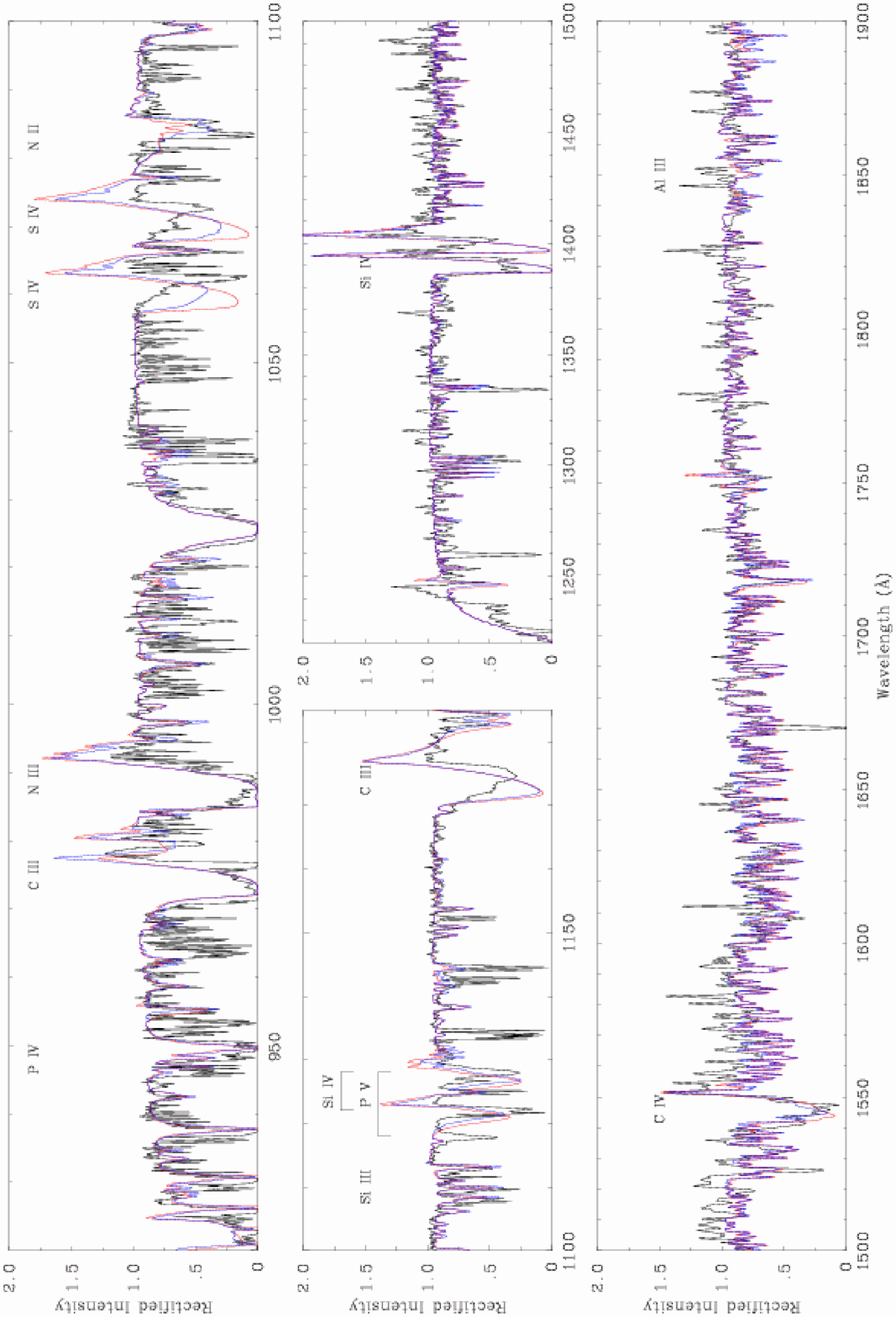}
\caption{Comparison between $FUSE$/$IUE$ observations of AV\,235 and 
the H$\gamma$ derived {\sc cmfgen} model spectra (unclumped -- red; clumped -- blue).
Again the model spectra have been multiplied by an appropriate transmission spectrum to include
the effects of neutral hydrogen absorption.\label{av235_uv2}}
\end{center}
\end{figure*}

A consequence of the hotter model is that many of the UV features are
in better agreement with the observations, namely P~\4, N~\2, Si~\3,
C~\4 and Al~\3 as shown in Figure~\ref{av235_uv2}.  This solution
appears to be more representative for the true metal ionization
balance, although the theoretical S~\4 and Si~\4 profiles are still strongly
overpredicted; clumped models help slightly but do not resolve the problems.

Both the intensity and morphology of the H$\alpha$ emission agree in
our VLT-UVES and ANU spectra.  Additionally, the feature is identical
in a CASPEC echelle spectrum taken using the ESO 3.6-m telescope in
Oct.  1991 (Dr.~Daniel Lennon, private communication), i.e., there is no obvious
evidence for significant H$\alpha$ variability.  Thus, clumping aside,
the H$\gamma$/far-UV and H$\alpha$ wind diagnostics indicate a
discrepancy in mass-loss of at least a factor of two.  Ultimately, all
optical diagnostics for AV\,235 can be reproduced in the case of a
clumped, weak wind at \Teff~=27.5kK, with the exception of H$\alpha$. 
It would seem that in this case at least our assumption of spherical 
symmetry is not valid.

\section{CNO abundances}
\label{abundances}

Derived CNO abundances (by number) for our present sample are given
in Table~\ref{results_cno}, together with 
those from Paper~I and Hillier et al. (2003). We also include
H\,{\sc ii} region abundances from Russell \& Dopita (1992), plus
more recent stellar abundances in the SMC from Venn (1999).

AV\,488 is in common with the quantitative analysis of Lennon et al.
(1991), for which a lower temperature of 25\,kK was obtained. Their
photospheric analysis yielded CNO abundances of log(C/H)+12\,=\,7.3,
log(N/H)+12\,=\,7.7 and log(O/H)\,=\,7.7, within a factor of
three of the present values.

In Table \ref{results_cno} we also present new results for
Sk$-$66$^\circ$169.  In the course of this work it became apparent that
the source of the He~\2 \lam4686 discrepancy for this target in Paper 1 was
due mainly to the large adopted turbulent velocity in the model atmosphere
calculation, to which \lam4686 is very sensitive.  This discovery led to
revisions in the adopted parameters of Sk$-$66$^\circ$169 (a hotter
model of 27.5\,kK is now preferred), which in turn affects the derived
abundances.  This now enables us to make a fully consistent comparison
between the two O9.7-type stars.  

Walborn (1976) suggested that it is the OBC-type stars that are least
evolved, rather than the morphologically normal and OBN-types.  We
would therefore expect normal supergiants to show evidence for partial
CNO processing.  In general (e.g., Paper~I) ``normal'' O and B-type
supergiants exhibit evidence of nitrogen enrichment, together with
carbon and oxygen depletion.  In the current sample, this is
especially evident for HDE\,269896 in which C/N=0.008, versus
C/N=0.075 in Sk$-$66$^\circ$169 (and C/N=8 for the ISM),
i.e., HDE\,269698 appears to be more fully processed.  In contrast, the
only OBC-type star from the combined sample, AV\,69 (Hillier et al. 2003)
reveals normal SMC C/N abundances. Given that AV\,69 is an O7.5 giant,
we now discuss an OBC supergiant that has stellar parameters much
closer to the present sample to verify our claims, with a particular
emphasis on the nitrogen abundance.

BC-type supergiants are rare and so we resort to an abundance analysis of
the Galactic BC0.7~Ia HD~2905 (Lennon et al. 1993), using identical
techniques to those employed for the present sample.  Photometric and
distance information was taken from Humphreys (1978), i.e., $V = 4.16,
B-V = 0.14$ with a distance modulus of 10.2 mag (i.e., 1.1 kpc)
appropriate for Cas~OB14. Observational data are drawn from Smartt et
al. (2002) for which HD~2905 served as a Galactic early B abundance
standard. The absolute visual magnitude of $M_{\rm V}=-$7.4 mag
follows from the final model intrinsic colour, $(B-V)_0=-$0.27.

\begin{figure}
\begin{center}
\includegraphics[scale=0.80]{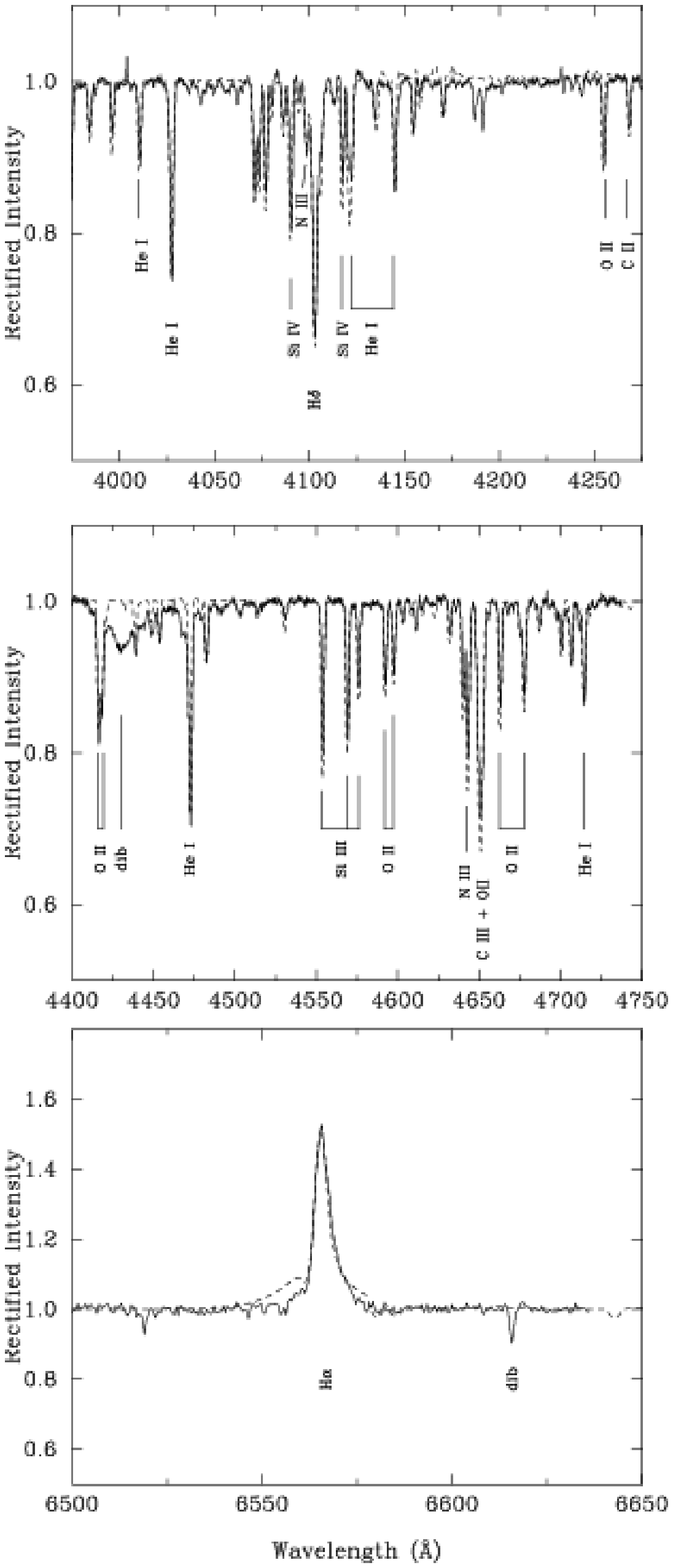}
\caption{Comparison of the HD~2905 optical data with the model fit (dashed line) in
the two primary diagnostic blue regions and in the vicinity of H$\alpha$.  From
blue to red wavelengths, by species, the labelled lines are: He~\1 \lam\lam4009, 4026, 
4116, 4121, 4144, 4471, 4713; C~\2 \lam4267; N~\3 \lam\lam4097, 4640; O~\2 \lam\lam4255, 4415-17, 
4591-96, 4661-73-76; Si~\3 \lam4553-68-75; Si~\4 \lam\lam4089, 4116 and the C~\3$+$O~\2
\lam4650 blend.  Diffuse interstellar bands are also marked at \lam\lam4430 and 6614.\label{2905_fig}}
\end{center}
\end{figure}

In the absence of $FUSE$ far-UV spectroscopy for HD~2905, we adopt
$v_\infty = 1105$~\kms\ for our analysis, plus $v \sin i$=91
\kms~(Howarth et al. 1997). Our derived parameters for HD~2905 are
given in Table \ref{2905res} together with recent non-LTE results by
Kudritzki et al. (1999) and Smartt et al. (2002). Comparisons between
the final model and the observed spectrum are shown in
Figure~\ref{2905_fig}.  For consistency with the other early B-type
supergiants in our sample we adopted a microturbulence of $\xi =
20$\kms.  This value gives excellent agreement for the He~\1/\2 and
Si~\3 lines however, as also seen in AV\,70 and 372, the predicted
intensity of the Si~\4 \lam4116 line is too strong.

Our derived temperature, $T_{\rm eff}$= 22.5 kK, is 1000-1500\,K lower
than found by Kudritzki et al. or Smartt et al. The former study
merely adopted the appropriate Si~\4/Si~\3 temperature scale from the
McErlean et al. (1999) non-LTE unblanketed, plane-parallel work,
whilst an equivalent analysis was carried out by Smartt et al. (2002).
Using the approximate methods of Puls et al. (1996), the mass-loss
rate determined for HD\,2905 by Kudritzki et al. (1999) is in
reasonable agreement with the present study.

The diagnostic nitrogen lines (i.e., N~\2 \lam4601--43, 3995) are well
matched by a 1.3 times Solar abundance of log(N/H)+12=8.15.  The
numerous O~\2 features are best fit with an approximately Solar oxygen
abundance of log(O/H)+12=8.7, which closely matches the region in the
vicinity of $\lambda$4650. Carbon is more problematic, such that we
tentatively adopt a 1/5 Solar abundance of log(C/H)+12$\sim$8 from
C~\2 \lam4267.

Therefore, HD~2905 reveals marginally processed CNO
abundances. Assuming it was formed from moderately sub-solar ISM
material, carbon has been slightly reduced with nitrogen showing a
modest enrichment, and oxygen unaffected.  It would certainly be of
interest to revisit further Galactic targets, e.g., those studied by
Massa et al. (1991).  They found evidence of significant nitrogen
enhancement in the Galactic BN1 star HD\,93840, yet solar
abundances for the normal comparison star, $\zeta$ Per.  Their results
for the morphologically normal star are not necessarily at odds with
our values for HD~2905 since both of their targets are
less luminous type-Ib supergiants, for which mass-loss may have had
less impact on the appearance of the atmospheres.

From our analysis of HD~2095 we take greater confidence in our present
results, such that the general nitrogen enrichment versus the
Magellanic Cloud ISM abundances appears to be genuine. For comparison,
Smartt et al. obtained significantly smaller values (over a factor of
five) using non-LTE, plane-parallel (unblanketed) models for their
analysis.  This indicates the degree of uncertainty in the absolute
abundance ratios for B supergiant analyses when different techniques
are used.

The implications of the inclusion of rotation on surface abundances
from theoretical evolutionary models are well documented
e.g., Meynet (1998), Heger \& Langer (2000) and Maeder \& Meynet (2001).
Qualitative evidence of rotationally-induced mixing was given by Howarth 
\& Smith (2001) who found that ON-type main sequence stars were
drawn from a more rapidly rotating population than those with
morphologically normal spectra.  More recently, the enhanced nitrogen
abundances found in Paper I and in AV\,83 (Hillier et al, 2003) were
attributed to rotational mixing, in combination with the effects of CNO
processing.

We obtain significantly enhanced nitrogen abundances for all of our
targets.  When compared to the abundances from Rusell \& Dopita
(1992), the N/C ratios for the current sample represent enhancements
by factors of about 100.  Enhanced nitrogen abundances were found by Maeder
\& Meynet (2001) in their high-mass, fast-rotating (\vsini~=~300 \kms)
models, however it seems unlikely that our results are solely
attributable to rapid rotation.  Indeed if high rotational velocities
were implicated, our targets would have ``spun down'' more quickly
than the Maeder \& Meynet models (see their Table~1).  A more
plausible solution is that rotational mixing is more effective than
previously thought, even at relatively moderate velocities (as also
concluded by Trundle et al. 2004).

In Table~\ref{results_cno} we also include abundances derived for the
nebula of the luminous blue variable (LBV) R127 in the LMC. It is
clear that the stellar enrichment as indicated by N/O of some OB
supergiants actually exceed those of the LBV nebula.

\section{Temperature calibrations for OB supergiants}

Several recent studies have indicated that commonly adopted
temperatures for O-type stars were too high. Martins et al. (2002)
used line-blanketed, spherical models for Galactic O-type dwarfs to
indicate a modest downward revision, whilst a substantial downward
revision was indicated for extreme O-type supergiants in Paper~I using
similar techniques.  In their analyses of Galactic O-type stars both
Herrero et al. (2002) and Bianchi \& Garcia (2002) also found lower
temperatures than from previous studies (although the methods of
Bianchi \& Garcia differ in that they rely solely on the UV-region,
neglecting the traditional optical lines).  Such downward revisions of
temperatures (of order 10-20$\%$) are commensurate with the initial
line-blanketed results of Hubeny et al.  (1998) for 10 Lac.

\begin{figure}
\begin{center}
\includegraphics[scale=0.80]{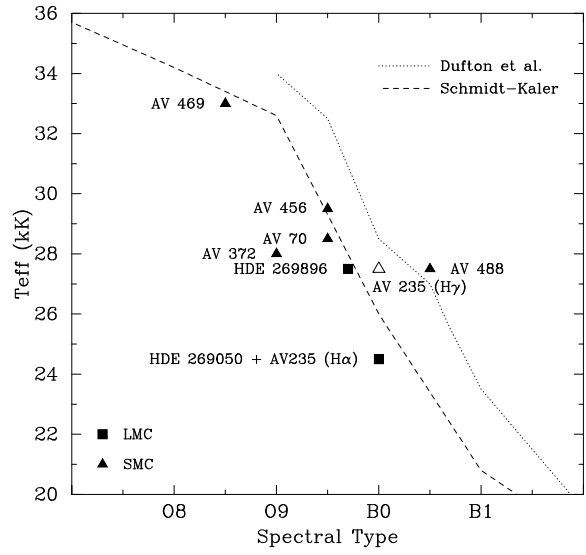}
\caption[]{Comparison of derived temperatures with selected calibrations
for our current targets in the LMC (squares) and SMC (triangles).
Also shown are published temperature scales from Schmidt-Kaler
(1982; dashed line) and Dufton et al. (2000; dotted line).\label{teff}}
\end{center}
\end{figure}

Our derived temperatures are plotted as a function of spectral type in
Figure \ref{teff}, together with the widely used Schmidt-Kaler (1982)
calibration and that of Dufton et al. (2000), which represents the
recently adopted late O and early B-type calibration based on
unblanketed {\sc tlusty} models (McErlean et al. 1999).  This
neatly illustrates the effect played by blanketing and stellar
winds when compared to the unblanketed temperatures; in general we
find differences of $\sim$2\,kK.  However, AV\,488 deviates from this
general trend in the sense that it lies perfectly on the unblanketed
results.  This result is consistent with the temperatures found from
recent line-blanketed, non-LTE analyses of early B-type SMC
supergiants by Trundle et al. (2004).  As Table~\ref{table-teff}
indicates, temperatures of stars with strong winds (H$\alpha$ in
emission), deviate more from standard (plane-parallel) calibrations
than those with weak winds (H$\alpha$ in absorption) such as AV\,488,
as might be anticipated. 

Published temperature calibrations have always been monotonic in
nature, i.e., \Teff~decreases with later spectral types, (e.g., Vacca et
al. 1996), but this is only the case when one separates extreme supergiants
(typically Ia$^{+}$ luminosity class) from normal supergiants
(typically Ib, II).  Therefore, considering the H$\alpha$ model for
the extreme B0\,Ia supergiant AV\,235 together with AV\,488 (B0.5~Ia),
it is the high wind density of the former and low wind density of the
latter which conspire to upset the conventional downward sequence.
The peculiarity of this situation is reduced when considering the
H$\gamma$ model; a hotter temperature is required to match the optical
spectrum when a lower wind density (i.e., smaller mass-loss rate) is required.
Similar arguments have already been made recently by Herrero et
al. (2002). 

Further work on a yet larger Magellanic Cloud sample including $FUSE$
and UVES observations, spanning early-type dwarfs and later B-type
supergiants is ongoing and should further elucidate the temperature
scale for early-type stars.  Similarly, a larger study of Galactic
early B supergiants is currently in progress which confirms the
present temperature scale for B supergiants with or without strong
winds (Crowther et al. 2004, in preparation).

\section{Wind density and momentum} 
The wind-momentum-luminosity relationship (WLR, Puls et al. 1996),
relating \mdot, $v_\infty$ and $(R_{\ast} / R_{\odot})^{0.5}$ to stellar
luminosity has been claimed to provide a means by which distances to
galaxies containing OBA-type supergiants may be obtained (e.g.,
Kudritzki et al. 1999). Indeed, in external galaxies B-type
supergiants are generally considered to be more useful than O-types as
they are visually brighter.  There are two direct consequences of the
present results (together with those from Paper~I) regarding the
calibration of the WLR.  Firstly, as a direct consequence of lower
temperatures from our blanketed model atmospheres, derived
luminosities are lower (recall Figure~19 from Paper~I).  Also, given
the indications that OB-type winds are clumped, there will be a second
correction to the absolute value of the wind-momentum.  Unfortunately,
determining clumping factors remains a formidable challenge.

\begin{figure}
\begin{center}
\includegraphics[scale=0.80]{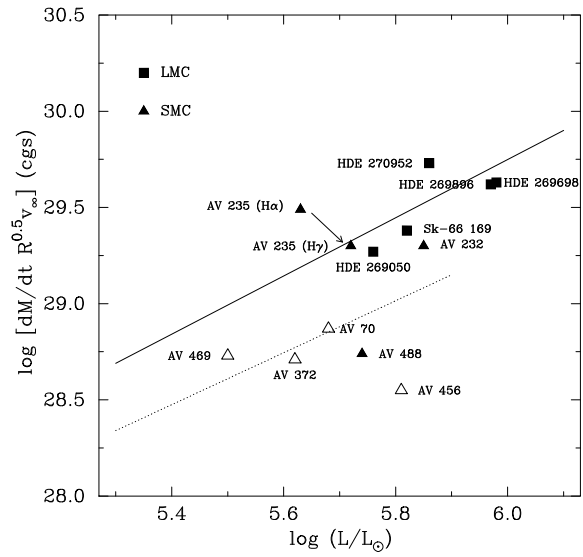}
\caption{Reduced wind momentum as a function of luminosity for both the current
sample and the four stars from Paper I.  The type Ia stars are shown
as solid symbols, the less luminous stars as open symbols.  The
form of the wind-momentum-luminosity relationship for Galactic O and
early B-type supergiants (solid and dotted lines respectively) from
Kudritzki \& Puls (2000) are shown.  Two values are shown for AV\,235, 
based on (homogeneous) mass-loss rates obtained from H$\alpha$ and the H$\gamma$ (see
\sref{av235}).\label{wlr}}
\end{center}
\end{figure}

Kudritzki et al. (1999) discuss differences in wind driving lines
between O, B and A-type supergiants, such that each will possess different
scaling laws (see also Kudritzki \& Puls 2000).  We therefore present
our current (unclumped) results in Figure~\ref{wlr} together with
calibrations from (unblanketed) models of Galactic OB-type
supergiants obtained by Kudritzki \& Puls. We present two values for
AV\,235 given the contradictory mass-loss diagnostics for this
particular star. Herrero et al. (2002) provide an updated calibration
for Galactic O-type supergiants using blanketed model results, with a
similar slope to Kudritzki \& Puls for high luminosities, but
$\sim$0.2 dex lower for moderate luminosities.

The Magellanic Cloud stars from our present sample generally lie
within a factor of two of the Kudritzki \& Puls calibration, in spite
of the lower luminosities as a result of blanketing and (predicted)
weaker winds as a result of lower metallicity. At first glance this is
rather puzzling.  However, we include in our sample the most extreme
OB-type supergiants in the Magellanic Clouds, such that they will
provide rather poor templates with which Galactic supergiants should
be compared. Consequently, firm results will only be possible once
large numbers of O and B-type stars have been studied in the Milky
Way, LMC and SMC. Work towards this goal is presently underway by
various groups. Clumping, with volume filling factors of order 10\%,
in which mass-loss rates are actually a factor of three times lower,
would serve to reduce the calibration by 0.5 dex.

\begin{figure}
\begin{center}
\includegraphics[scale=0.80]{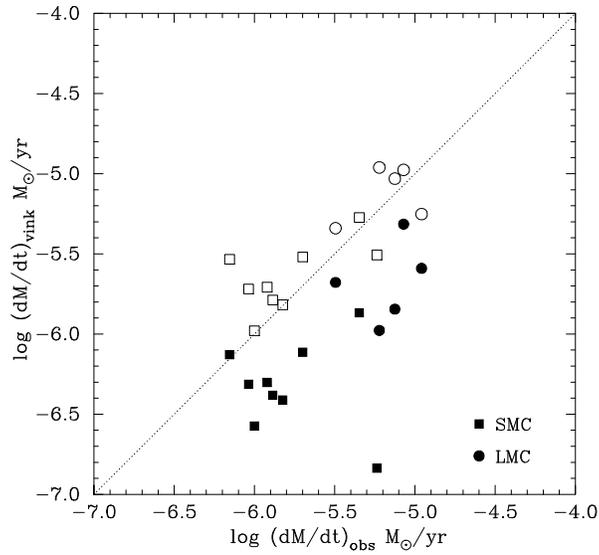}
\caption{Comparison between theoretical (Vink et al., 2001) and 
observed (unclumped) mass-loss rates for our sample.  Open symbols are
theoretical values calculated for $Z/Z_\odot = 1$, while solid symbols
indicate computed values for LMC ($Z/Z_\odot = 0.4$; circles) and SMC
($Z/Z_\odot = 0.2$; squares) metallicities.  The dotted line
highlights the 1:1 relationship.\label{vink}}
\end{center}
\end{figure}

The (unclumped) mass-loss rates for our sample are compared with the 
theoretical
predictions for our stars from the recipes of Vink et al. (2001) in Figure
\ref{vink}; since the methods are identical, results from 
Paper~I and Hillier et al. (2003; using the unclumped $\dot{M}$ for
AV\,83) are also included.  In the figure the open symbols are the
theoretical mass-loss rates calculated for $(Z/Z_\odot) = 1$.  The
solid symbols are the theoretical mass-loss rates calculated for the
appropriate metallicity, i.e., $(Z/Z_\odot)$ = 0.2 or 0.4; the diagonal
dotted line simply indicates the 1:1 relationship.  The plot is
somewhat complicated by the fact for some stars there is not a
constant offset between points for the same star at the two
metallicities. This arises because some of our stars (most notably
AV\,235, with the lowest predicted mass-loss rate) lie between the two
bi-stability jumps.  In these cases the jump is recalculated by Vink's
routine, which due to the metallicity dependence (see Vink et al., Eqn.~15),
leads to different coefficients when calculating the predicted
rates; similar problems were encountered by Trundle et al. (2004).

The first point to note from the figure is that the observationally
derived mass-loss rates are generally higher for our LMC stars than
those in the SMC; this is simply a selection effect of the current
overall sample.  Four of the five LMC stars are classified as Ia$^+$,
whereas the SMC stars are generally less extreme.  This is unfortunate
as it makes it difficult to perform meaningful comparisons between the
two metallicities.  A second comment is that the theoretical results
at $(Z/Z_\odot) = 1$ (i.e., assuming no $Z$ dependence) are
qualitatively a better match to the observed mass-loss rates than
those taking metallicity into account.  Similarly, for three early
B-type supergiants in the SMC, Trundle et al. (2004) also found that
the observationally derived mass-loss rates were larger than those
predicted for $(Z/Z_\odot)$ = 0.2.  Vink et al. used different sets of
stellar atmosphere models and thus no physical significance should be
attached to these results at the current time.  They do however
reinforce the need for further theoretical efforts in the sense that
the ``recipe'' to predict stellar-mass loss rates gives different
values to the analyses here and to those by Trundle et al. (2004).
Similarly, if one is to reliably test the dependence of stellar
mass-loss rates with metallicity, it is clear that we require a large,
$homogenous$ observational study of early-type stars in the Clouds.

\section{Conclusions}
\label{end}

We have studied a sample of LMC and SMC late O and early B supergiants
based on modern line-blanketed, spherical models plus extensive $FUSE$
(far-UV), $IUE$ (UV) and UVES (visual) observations. In general, we
find excellent agreement between alternative optical temperature
diagnostics (e.g., He and Si) with H$\alpha$ derived mass-loss rates.
Adopting homogeneous models, some UV wind features are systematically
too strong.

The UV discrepancies are reduced if OB winds are clumped. P\,{\sc v}
$\lambda\lambda$1118-1128 was identified as a useful probe of clumping
in O supergiants in Paper~I, unless phosphorus is depleted relative to
other elements in the Magellanic Clouds.  In the present study, S\,\4
$\lambda\lambda$1063-1072 appears to offer an equivalent probe in
early B-type supergiants, with the additional benefit that the ISM
sulphur abundance is well known (Russell \& Dopita 1992). We conclude
that winds in OB-type stars are at least moderately clumped.  This leads
to lower derived mass-loss rates than otherwise, which scale with the adopted
filling-factor, (with $f_\infty = 0.1$, $\dot{M}$ is reduced by a
factor of $\sim$3).  AV\,235 (B0\,Iaw) is peculiar in the current sample
in that it has inconsistent optical (Balmer line), UV and far-UV wind
features. All diagnostics indicate a relatively weak (and clumped)
wind for AV\,235 with the exception of H$\alpha$.

Stellar temperatures for O and early B-type supergiants are generally
2--4\,kK lower than recent calibrations based on unblanketed,
plane-parallel models. Supergiants with extreme (typically Ia$^{+}$)
winds are more greatly affected, such that there is expected to be
different spectral type--temperature calibrations for OB supergiants,
depending on whether H$\alpha$ is in emission. Reduced temperatures
consequently indicate lower luminosities, with impact upon ionizing
fluxes (recall Figure~13 from Herrero et al. 2002), and wind-momentum
luminosity calibrations.

We also investigate CNO abundances for OB supergiants. As in Paper~I,
``normal'' OB supergiants are found to have partially processed
abundances, i.e., log(N/C)$\sim$1 versus log(N/C)$\sim-1$ for H\,{\sc
ii} regions.  Although such large nitrogen enrichments were found in
the fast-rotating evolutionary models from Maeder \& Meynet (2001), it
seems unlikely that rapid rotation is solely responsible and that
mixing is perhaps more effective than previously thought at more
moderate velocities.  A differential analysis of HDE\,269896
(ON9.7\,Ia$^{+}$) versus Sk$-66^{\circ}$169 (O9.7\,Ia$^{+}$) indicates
even more extreme abundances -- nitrogen is further enriched at the
expense of carbon and oxygen in the ON supergiant. In contrast, two
OBC stars exhibit fairly normal CNO abundances; HD~2905 studied here
for comparison to the Magellanic Cloud B-type supergiants, plus AV\,69
investigated by Hillier et al. (2003) using identical techniques.
Consequently, quantitative models now provide strong evidence for the
sequence OBC (normal CNO) $\longrightarrow$ OB (CNO partially
processed) $\longrightarrow$ OBN (fully CNO processed) originally
suggested on from morphological arguments by Walborn (1976, 1988) and
supported by studies of He-contents by Smith \& Howarth (1994).

\section{Acknowledgements}
Financial support has been provided by PPARC (CJE), the Royal Society
(PAC).  JDH acknowledges support from NASA grants NAGW-3828 and
NAG5-10377 and NASA Space Telescope Science Institute (STScI) grant
AR07985.02-96A.  We thank Lex Kaper for providing his UVES data,
Thierry Lanz for supplying a {\sc tlusty} B-supergiant model, Stephen
Smartt for kindly providing the HD2905 data and Jay Abbott for
reducing the ANU data.

\clearpage

\begin{center}
\begin{deluxetable}{llclccccccccc}
\rotate
\tabletypesize{\footnotesize}
\tablewidth{0pc}
\tablecolumns{13}
\tablecaption{Observational parameters of target stars \label{targets1}}
\tablehead{
\colhead{Star} & \colhead{Alias} & \colhead{Galaxy} & \colhead{Sp. Type} & \colhead{Ref.} & 
\colhead{$V$} & \colhead{$B - V$} & \colhead{Ref.} & \colhead{$E(B - V)$} & 
\colhead{$M_V$} & \colhead{v$_r$} & \colhead{log $N$(H\,\1)} & \colhead{log $N$(H$_2$)} \\
 & & & & & \colhead{mag} & \colhead{mag} & & \colhead{mag} & \colhead{mag} & 
\colhead{[\kms]} & \colhead{[cm$^{-2}$]} & \colhead{[cm$^{-2}$]}
}
\startdata
AV\,469 & Sk 148 & SMC & O8.5 II((f)) & 7 & 13.20 & $-$0.22 & 2 & 0.09 & $-$6.0 & 187 & 21.3& --\\
AV\,372 & Sk 116 & SMC & O9 Iabw & 7 & 12.59 & $-$0.15 & 4 & 0.12 & $-$6.7 & 262 & 21.8& 16.5\\
AV\,70 & Sk 35 & SMC & O9.5 Ibw & 6 & 12.38 & $-$0.17 & 2 & 0.10 & $-$6.8 & 170 & 21.3 & --\\
AV\,456 & Sk 143 & SMC & O9.5 Ibw & 8 & 12.83 & $\phantom{-}$0.10 & 4 & 0.35 & $-$7.2 & 169 & -- & --\\
HDE\,269896 & Sk-68$^\circ$135 & LMC & ON9.7 Ia$^+$ & 5 & 11.36 & $\phantom{-}$0.00 & 1 & 0.25 & $-$7.9 & 290 &21.5& 19.9\\
HDE\,269050 & Sk-68$^\circ$52 & LMC & B0 Ia & 5 & 11.54 & $-$0.07 & 1 & 0.17 & $-$7.5 & 234 & 21.6 & 19.5\\
AV\,235 & Sk 82 & SMC & B0 Iaw & 5 & 12.20 & $-$0.18 & 2 & 0.07 & $-$6.9 & 161  & 21.3 & 15.9 \\
AV\,488 & Sk 159 & SMC & B0.5 Iaw & 6 & 11.90 & $-$0.13 & 3 & 0.09 & $-$7.3 & 200 & 18.9 & 18.9 \\
\enddata
\tablecomments{Reddenings and absolute magnitudes were derived from the mean of intrinsic colours from
Fitzgerald (1970) and from fitting stellar models to UV-optical spectrophotometry.  Neutral hydrogen 
column densities for our targets (with the exception of AV\,456) are determined from fits to the wings
of the Lyman $\beta$ line in the $FUSE$ spectra.  Molecular hydrogen column densities from Tumlinson et al. (2002)
are also given where available.}
\tablerefs{(1) Ardeberg et al. (1972), (2) Azzopardi \& Vigneau (1975), (3) Dachs (1970), (4) Massey (2002), 
(5) Walborn (1977), (6) Walborn (1983), (7) Walborn et al. (2002), (8) this work.}
\end{deluxetable}
\end{center}

\clearpage

\begin{center}
\begin{deluxetable}{llccccllc}
\rotate
\tabletypesize{\footnotesize}
\tablewidth{0pc}
\tablecolumns{9}
\tablecaption{Details of VLT-UVES, $FUSE$ and other UV observations of target stars \label{targets2}}
\tablehead{
\colhead{Star}  & \colhead{$FUSE$ ID} & \multicolumn{2}{c}{Exp. [min]} & \multicolumn{2}{c}{Date} & 
\colhead{Optical coverage [\AA]} & \colhead{Archival data} & \colhead{$IUE$ Image} \\
 & & \colhead{$FUSE$} & \colhead{VLT} & \colhead{$FUSE$} & \colhead{VLT} & & & 
}
\startdata
AV\,469 &  P1176301 & 136 & 33 & 2000/10/11 & 2001/09/27 & 3770--4945, 6380--8200 & $HST$-FOS & -- \\ 
AV\,372 &  P1176501 & $\phantom{1}$73 & 40 & 2000/10/11 & 2001/09/27 & 3770--4945, 6380--8200 & $HST$-FOS & -- \\
AV\,70 &  B0900601 & 103 & 28 & 2001/06/15 & 2001/09/24 & 3400--5750, 5830--6800 & $IUE$-LORES & SWP18830 \\
AV\,456 &  Q1070101 & $\phantom{1}$62 & 30 & 2000/10/11 & 2001/09/24 & 3400--4500, 4670--6650 & $IUE$-LORES & SWP45199 \\
HDE\,269896 & P1173901 & 118 & 29 & 2000/02/12 & 2001/09/24 & 3400--5750, 5830--8310 & $IUE$-HIRES & SWP47594 \\
HDE\,269050 & P1174001 & 135 & 20 & 2000/10/01 & 2001/09/29 & 3770--4945, 6380--8200 & $IUE$-HIRES & SWP53054 \\
AV\,235 &  P1030301 & 270 & 17 & 2000/07/02 & 2000/09/27 & 3770--4945, 6380--8200 & $IUE$-HIRES & SWP13535\\
AV\,488 &  P1030501 & 113 & 28 & 2000/10/04 & 2001/09/24 & 3400--5750, 5830--8510 & $IUE$-HIRES & SWP16614\\
\enddata
\end{deluxetable}
\end{center}

\clearpage

\begin{center}
\begin{deluxetable}{llccccccccc}
\rotate
\tabletypesize{\scriptsize}
\tablewidth{0pc}
\tablecolumns{11}
\tablecaption{Derived parameters of target stars \label{results}}
\tablehead{
\colhead{Star} & \colhead{Sp. Type} & \colhead{\Teff\tablenotemark{a}} & \colhead{$R_\ast$} & \colhead{\logg} & 
\colhead{log($L$/$L_\odot$)} & \colhead{$\dot{M}$} & \colhead{$f$} & \colhead{$\beta$} & \colhead{$v_\infty$} & \colhead{\vsini} \\
& & \colhead{[kK]} & \colhead{[$R_\odot$]} & \colhead{[cgs]} & & \colhead{[$M_\odot$ yr$^{-1}$]} & & & \colhead{[\kms]} & \colhead{[\kms]}
} 
\startdata
AV\,469 & O8.5 II((f)) & 33.0 & 17.2 & 3.4 & 5.50 & $\phantom{1}$1.3$\times$10$^{-6}$ & 1.0 & 1.00 & 1550 & $\phantom{1}$80 \\ 
& & & & & &$\phantom{1}$ 5.0$\times$10$^{-7}$ & 0.1 & & & \\

AV\,372 & O9 Iabw & 28.0 & 27.5 & 3.1 & 5.62 & 1.0$\times$10$^{-6}$ & 1.0 & 2.25 & 1550 & 120 \\
& & & & & & $\phantom{1}$3.5$\times$10$^{-7}$ & 0.1 & 2.75 & & \\

AV\,70 & O9.5 Ibw & 28.5 & 28.4 & 3.1 & 5.68 & $\phantom{1}$1.5$\times$10$^{-6}$ & 1.0 & 1.75 & 1450 & 100 \\ 
& & & & & & $\phantom{1}$4.5$\times$10$^{-7}$ & 0.1 & & \\

AV\,456 & O9.5 Ibw & 29.5 & 30.6 & 3.0 & 5.81 & $\phantom{1}$7.0$\times$10$^{-7}$ & 1.0 & 1.75 & 1450 & $\phantom{1}$80 \\
& & 30.5 & 28.6 & & & $\phantom{1}$3.0$\times$10$^{-7}$ & 0.1 & & & \\

HDE\,269896 & ON9.7 Ia$^+$ & 27.5 & 42.3 & 2.7 & 5.97 & $\phantom{1}$7.5$\times$10$^{-6}$ & 1.0 & 3.50 & 1350 & $\phantom{1}$70 \\ 
& & & & & & $\phantom{1}$2.5$\times$10$^{-6}$ & 0.1 & & &  \\

HDE\,269050 & B0 Ia & 24.5 & 42.2 & 2.7 & 5.76 & $\phantom{1}$3.2$\times$10$^{-6}$ & 1.0 & 2.75 & 1400 & $\phantom{1}$80 \\
& & 25.5 & 39.0 & & & $\phantom{1}$9.0$\times$10$^{-7}$ & 0.1 & & & \\

AV\,235 (H$\alpha$) & B0 Iaw & 24.5 & 36.2 & 2.8 & 5.63 & $\phantom{1}$5.8$\times$10$^{-6}$ & 1.0 & 2.50 & 1400 & $\phantom{1}$80 \\ 
& & & & & & $\phantom{1}$1.5$\times$10$^{-6}$ & 0.1 & & & \\

AV\,235 (H$\gamma$) & & 27.5 & 31.9 & 2.9 & 5.72 & $\phantom{1}$4.0$\times$10$^{-6}$ & 1.0 & 1.50 & 1400 & $\phantom{1}$80 \\ 
& &  &  & & & $\phantom{1}$1.3$\times$10$^{-6}$ & 0.1 & & & \\

AV\,488 & B0.5 Iaw & 27.5 & 32.6 & 2.9 & 5.74 & $\phantom{1}$1.2$\times$10$^{-6}$ & 1.0 & 1.75 & 1250 & $\phantom{1}$80 \\
& & & & & & $\phantom{1}$4.5$\times$10$^{-7}$ & 0.1 & & & \\
\enddata
\tablenotetext{a}{As is usual for stars with extended atmospheres, stellar temperatures are defined relative to a 
radius of Rosseland optical depth 10.}
\tablecomments{Mass-loss rates are given for homogeneous winds (with a maximum volume filling factor $f$=1) and
for clumped winds ($f$=0.1); in two instances the introduction of clumping necessitated slight changes in 
the other parameters to fit H$\alpha$ and the other optical lines successfully.  Two sets of results are given 
for AV\,235 from fits to the H$\alpha$ and H$\gamma$ lines (see \sref{av235}).}
\end{deluxetable}
\end{center}

\clearpage

\begin{center}
\begin{deluxetable}{llccccccr}
\rotate
\tabletypesize{\scriptsize}
\tablewidth{0pc}
\tablecolumns{9}
\tablecaption{Derived CNO abundances of target stars, including those from Paper~I and
Hillier et al. (2003) \label{results_cno}}
\tablehead{
\colhead{Star} & \colhead{Sp Type} & \colhead{$\epsilon$(He)} & \colhead{log(C/H)+12} & 
\colhead{log(N/H)+12} & \colhead{log(O/H)+12} & \colhead{log (N/C)} & \colhead{log (N/O)} & \colhead{Ref}
}
\startdata
& \multicolumn{5}{c}{Sun} \\
           &           & 0.09& 8.51 & 7.93 & 8.66 & $-$0.6\pd & $-$0.7\pd \\ 
& \multicolumn{5}{c}{LMC}\\
HDE\,269698 & O4\,Iaf$^+$  & \p0.2:& 7.3 & 9.1 & 7.8  & 1.8 & 1.3 &1\\
HDE\,270952 & O6\,Iaf$^+$  & \p0.2:& 7.65  & 8.8 & 7.7& 1.1 & 1.1 & 1\\
Sk--66$^{\circ}$169 & O9.7\,Ia$^+$
                          & \p0.2: & 7.3 & 7.95 & 8.1 &0.65& $-$0.15\pd & 3 \\
HDE\,269896 & ON9.7\,Ia$^+$& 0.2 & 7.4 & 8.3 & 8.0 & 0.9 & 0.3 & 3\\
HDE\,269050 & B0\,Ia       & 0.2 & 7.8 & 8.6 & 8.5 & 0.8 & 0.1 & 3\\
\tableline
RD92        & H\,{\sc ii} & -- & 8.04 & 7.14 & 8.35 & $-$0.9\pd & $-$1.21\pd \\
R127        & LBV         & -- &   --   & 8.05& 8.10  &   --    & $-$0.05\pd & \\
\tableline
& \multicolumn{5}{c}{SMC}\\
AV\,232    & O7\,Iaf$^+$ & \p0.2: & 7.5 & 8.45 & 8.0 & 0.95 & 0.45 & 1		 \\
AV\,83     & O7\,Iaf$^+$ & 0.2 & 7.6 & 8.4 & 7.8   & 0.8  & 0.6  &2 \\
AV\,69     & OC7.5\,III(f)&0.1 & 7.6 & 6.3 & 8.2   & $-$0.7\pd & $-$1.9\pd & 2 \\
AV\,469    & O8.5II((f))& 0.2 & 7.1 & 8.2 & --    & 1.1  &  --   & 3\\
AV\,372    & O9\,Iaw    &0.15 & 7.3 & 7.8 & --   & 0.5 &  --   & 3\\
AV\,70     & O9.5\,Iaw  &\p0.2: & 7.1 & 8.0 & $<$8.0& 0.9  & $>$0  &3\\
AV\,456    & O9.5\,Iaw  & 0.1  & --  & 7.7 & $<$8.0 & --  & -- &3 \\
AV\,235 (H$\alpha$)    & B0\,Iaw    & 0.2  & 7.0 & 8.4 & 8.1 & 1.4  & 0.3  &3\\
AV\,235 (H$\gamma$)    & B0\,Iaw    & 0.2  & 7.3 & 8.0 & 8.0 & 0.7  & 0.0  &3\\
AV\,488                & B0.5\,Iaw  & 0.2  & 7.2 & 8.1 & 8.0  & 0.9  & 0.1  &3\\
\tableline
RD92 & H\,{\sc ii}     & --  & 7.73 & 6.63 & 8.03 & $-$1.1\pd & $-$1.4\pd &\\
Venn &                 & --  & 7.4  & 6.6  & 8.1 &  $-$0.8\pd &  $-$1.5\pd &\\
\enddata
\tablecomments{Also included are Solar abundances (Grevesse \& Sauval 1998; Asplund
2003), Magellanic Cloud H\,{\sc ii} region abundances (Russell \& Dopita 1992, Venn 1999) 
and that of the nebula of R127 in the LMC (Smith et al. 1998).  Results from the two AV\,235 
models are given (see \sref{av235}).}
\tablerefs{(1) Paper~I; (2) Hillier et al. (2003); (3) This work}
\end{deluxetable}
\end{center}

\clearpage

\begin{center}
\begin{deluxetable}{lp{2cm}p{2cm}p{2cm}}
\tabletypesize{\footnotesize}
\tablewidth{0pc}
\tablecolumns{4}
\tablecaption{Comparison of derived parameters for HD 2905 (BC\,0.7Ia) with previous studies \label{2905res}}
\tablehead{
\colhead{Parameter} & \colhead{K99\tablenotemark{a}} & \colhead{S02\tablenotemark{b}} & \colhead{This work}
}
\startdata
$T_{\rm eff}$ (kK)   & 24   & 23.5 & 22.5  \\
$R_\ast (R_{\odot}$) & 41   &       & 47.7  \\
\logg                &2.7   & 2.7   & 2.7 \\
log($L$/$L_\odot$)   & 5.7  &       & 5.72 \\ 
$\dot{M}$ ($M_\odot$ yr$^{-1}$)  & 2.3$\times$10$^{-6}$  & --      & 2.4$\times$10$^{-6}$ \\
$\beta$              & 1.35 &   ---      & 2 \\
$\zeta$ (\kms)       & 11    &  11    & 20 \\
\vsini\ (\kms)        &     & 80     & 91\tablenotemark{c} \\
$M_{\rm V}$          & $-$7.0 &        &  $-$7.4  \\
log(C/H)+12  & & 7.0:   & 8.0 \\
log(N/H)+12  & & 7.3    & 8.15 \\
log(O/H)+12  & & 9.1    & 8.7 \\ 
\enddata
\tablenotetext{a}{Kudritzki et al. (1999)}
\tablenotetext{b}{Smartt et al. (2002)}
\tablenotetext{c}{Howarth et al. (1997)}
\end{deluxetable}
\end{center}

\clearpage

\begin{center}
\begin{deluxetable}{llllllll}
\tabletypesize{\footnotesize}
\tablewidth{0pc}
\tablecolumns{8}
\tablecaption{Temperature calibrations for late O and early B supergiants \label{table-teff}}
\tablehead{
\colhead{Sp Type} & \colhead{SK82\tablenotemark{a}} & \colhead{HM84\tablenotemark{b}} & \colhead{L93\tablenotemark{c}} & 
\colhead{V96\tablenotemark{d}} & \colhead{D00\tablenotemark{e}} & \multicolumn{2}{c}{This work} \\
        &      &      &     &     &     & \colhead{H$\alpha$ em} & \colhead{H$\alpha$ abs}
}
\startdata
O8.5    & 33.4 & 33.0 &     &34.2  &      &       & 33.0 \\
O9      & 32.6 & 32.6 &32.0 & 32.7 & 34.0 &       & 28.0 \\
O9.5    & 29.3 & 29.9 &30.0 & 31.2 & 32.5 &       & 29.0 \\
O9.7    &      &      &27.5 &      &      & 27.5  &      \\
B0      & 26.0 & 28.6 &25.0 & 28.2 & 28.5 & 24.5: &      \\
B0.5    & 23.4 & 23.1 &22.0 &      & 27.0 &       & 27.5 \\
B0.7    &      &      &21.0 &      & 25.5 & 22.5  &      \\
B1      & 20.8 & 20.3 &20.0 &      & 23.5 &       &      \\
\enddata
\tablenotetext{a}{Schmidt-Kaler (1982)}
\tablenotetext{b}{Humphreys \& McElroy (1984)}
\tablenotetext{c}{\,Lennon et al. (1993)}
\tablenotetext{d}{Vacca et al. (1996)}
\tablenotetext{e}{Dufton et al. (2000)}
\end{deluxetable}
\end{center}

\clearpage


\clearpage

\clearpage

\clearpage

\clearpage

\clearpage

\clearpage

\clearpage

\clearpage

\clearpage

\clearpage

\clearpage

\clearpage






\end{document}